\documentclass[sn-mathphys,Numbered]{sn-jnl}

\usepackage{pifont}
\usepackage{graphicx}%
\usepackage{multirow}%
\usepackage{amsmath,amssymb,amsfonts}%
\usepackage{amsthm}%
\usepackage{mathrsfs}%
\usepackage[title]{appendix}%
\usepackage{xcolor}%
\usepackage{textcomp}%
\usepackage{manyfoot}%
\usepackage{booktabs}%
\usepackage{algorithm}%
\usepackage{algorithmicx}%
\usepackage{algpseudocode}%
\usepackage{listings}%
\usepackage{threeparttable}
\usepackage{multicol}
\usepackage{array}

\theoremstyle{thmstyleone}%
\theoremstyle{thmstyletwo}%

\theoremstyle{thmstylethree}%

\begin{document}

\title[Article Title]{Fragmented Layer Grouping in GUI Designs
Through Graph Learning Based on Multimodal
Information}

\author[1]{\fnm{Yunnong} \sur{Chen}}
\author[1]{\fnm{Shuhong} \sur{Xiao}}
\author[1]{\fnm{Jiazhi} \sur{Li}}
\author[3]{\fnm{Tingting} \sur{Zhou}}
\author[3]{\fnm{Yanfang} \sur{Chang}}
\author[3]{\fnm{Yankun} \sur{Zhen}}
\author[1,2]{\fnm{Lingyun} \sur{Sun}}
\author*[1,2]{\fnm{Liuqing} \sur{Chen}}\email{chenlq@zju.edu.cn}

\affil[1]{\orgdiv{College of Computer Science and Technology}, \orgname{Zhejiang University}, \city{Hangzhou}, \postcode{310027}, \state{Zhejiang}, \country{China}}

\affil[2]{\orgdiv{Alibaba-Zhejiang University Joint Research Institute of Frontier Technologies}, \city{Hangzhou}, \postcode{310027}, \state{Zhejiang}, \country{China}}

\affil[3]{\orgdiv{Alibaba Group}, \city{Hangzhou}, \postcode{311121}, \state{Zhejiang}, \country{China}}


\abstract{Automatically constructing GUI groups of different granularities constitutes a critical intelligent step towards automating GUI design and implementation tasks. Specifically, in the industrial GUI-to-code process, fragmented layers may decrease the readability and maintainability of generated code, which can be alleviated by grouping semantically consistent fragmented layers in the design prototypes. This study aims to propose a graph-learning-based approach to tackle the fragmented layer grouping problem according to multi-modal information in design prototypes. Our graph learning module consists of self-attention and graph neural network modules. By taking the multimodal fused representation of GUI layers as input, we innovatively group fragmented layers by classifying GUI layers and regressing the bounding boxes of the corresponding GUI components simultaneously. Experiments on two real-world datasets demonstrate that our model achieves state-of-the-art performance. A further user study is also conducted to validate that our approach can assist an intelligent downstream tool in generating more maintainable and readable front-end code.}




\maketitle

\begin{figure*}[htbp]
\centering 
\includegraphics[width=1\textwidth]{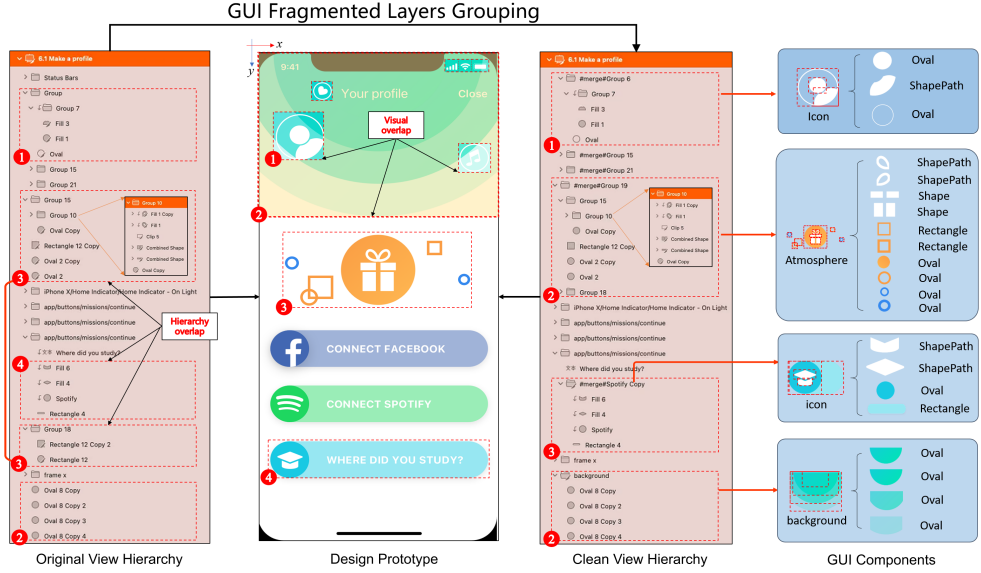} 
\caption{\textbf{Overview of Task Introduction and Challenge Specifications.} The design prototype contains a layer tree with a view hierarchy structure. UI layers are organized within this hierarchy, where layers of different components may overlap due to the aesthetic style of the design. By grouping fragmented layers, we can reorganize the GUI layers into a clean view hierarchy (as shown by \ding{172}\ding{173}\ding{174}\ding{175} in the figure). The four image blocks on the right visualize how each merged UI component is organized through layers.} 
\label{intro_img} 
\end{figure*}

\section{Introduction}\label{sec1}
Graphic User Interface (GUI) builds a visual bridge between software and end users. GUI design refers to the process of creating the interface layout, including the arrangement of visual elements like buttons, icons, and menus, as well as the interaction patterns that guide users through the software. A good GUI design makes the software efficient and easy to use, which has a significant influence on the success of applications and the loyalty of its users. In the industrial setting, the development of GUI starts from design prototypes produced with design software, such as Sketch \cite{sketchcli} and Figma \cite{figma}. A design prototype contains multiple GUI artboards. Each GUI artboard has a view hierarchy capturing the arrangement of GUI layers and it shows a visual effect of GUI design in the Canvas (as shown in Figure \ref{intro_img}). During the industrial GUI-to-code process, with a full understanding of the designer's intentions, front-end developers need to identify GUI components that should be instantiated on screen and rearrange them in a semantic structure to ensure correct visual displays \cite{moran2018machine}. To alleviate the burden on developers, some semi-automated code generation platforms (e.g. Imgcook \cite{imgcook}), which take design prototypes as inputs, are developed to generate code intelligently. In academia, intelligent code generation has also gained great attention of researchers \cite{beltramelli2018pix2code, GUI_skeleton, mohian2020doodle2app, xiao2024prototype2code}, and these methods are mainly based on GUI design images and cannot reach the great demand of industrial GUI development due to a lack of proper code structure and code accessibility.

At the beginning of industrial GUI development, grouping fragmented layers in GUI prototypes is a critical step towards intelligence to generate high-quality GUI code. Formally, fragmented layer grouping is defined as grouping layers that cannot independently convey visual semantics (referred to as fragmented layers) into GUI components (also called merging groups) to transform the disordered structure into a semantic structure in design prototypes. As depicted in Figure \ref{intro_img}, due to loose design standards, fragmented layers belonging to the same component are not grouped and are arranged by a disordered view structure (described as hierarchy overlap). During the process of transforming design prototypes to code by semi-automated tools, without grouping fragmented layers in a semantic structure, fragmented layers may be erroneously interpreted as separate entities. The misinterpretation can cause the generated code to contain redundant code snippets and correspond to a wrong GUI runtime hierarchy \cite{UILM,egfe}.

The fragmented layer grouping problem presents several challenges, which complicate the application of deep learning methods to this task. One of the primary obstacles, as shown in Figure \ref{intro_img}, is the disordered view hierarchy. This disorder is caused by nested groups and overlapping hierarchies, which forces us to abandon the original structural information. As a result, we must group layers from a flattened layer list, as highlighted in previous studies \cite{UILM,egfe}. Visual overlap is another challenge for grouping fragmented layers as it is hard to distinguish between background and foreground fragmented layers from GUI images. For example, there is a visual overlap between group \ding{172} and \ding{173} in Figure \ref{intro_img}. Last but not least, the number of fragmented layers and the number of merging groups vary across different design prototypes. Under such uncertainty, it is significant but challenging to discover all fragmented layers and determine the semantic consistency between them.

Previous studies have made attempts to address these challenges of fragmented layer grouping. They can be roughly summarized into two categories. Methods in the first category attempt to group fragmented layers based on GUI pixel images by computer vision algorithms \cite{UILM,UIED2}. To utilize the rich information contained in design prototypes, they create a semantic map by accessing the boundary information about layers and composite it with the GUI image to create a half-semantic image. For example, Chen et al. \cite{UILM} built up an object detection pipeline to detect the bounding boxes of merging groups based on the boundary prior. Then they find and merge fragmented layers inside each detected bounding box. Another category of methods adopts deep-learning techniques to directly discover semantic consistency between fragmented layers based on multimodal information. They assign labels to layers in design prototypes through sequence learning \cite{egfe} or graph learning \cite{li2022uldgnn} and group fragmented layers according to predicted labels. However, the previous methods above have some limitations. Object-detection-based approaches like UILM \cite{UILM} may falsely group fragmented and non-fragmented layers because they cannot distinguish them inside predicted bounding boxes. Layer-classification-based approaches like EGFE \cite{egfe} may falsely group fragmented layers of different merging groups due to hierarchy overlap. For example, in Figure \ref{intro_img}, EGFE \cite{egfe} may falsely merge fragmented layers in group \ding{175} and layers in the sub-group of \ding{174} due to its limitations.

In this study, we propose a new graph-learning algorithm to overcome the limitations of existing methods. To address the limitations, we innovatively detect the bounding boxes of merging groups and classify GUI layers inside the boxes based on a graph neural network. Specifically, we convert the irregular view hierarchy in a design prototype into a graph based on the inclusion relationship between GUI layers. In the process of graph learning, we introduce a self-attention module to graph learning blocks to overcome the over-smoothing problem. Compared with UILM \cite{UILM}, which only detects the bounding boxes of merging groups, our approach refines and updates feature vectors of GUI layers and then identifies fragmented layers inside the boxes. In this way, we can avoid the limitations of grouping non-fragmented layers falsely inside predicted bounding boxes. To overcome the limitations of EGFE \cite{egfe}, which is influenced by hierarchy overlap, we construct a new graph representation for design prototypes and adopt a graph neural network to better learn the complex GUI context. We conduct experiments on a real-world dataset to validate the effectiveness of our approach and propose four metrics to evaluate the performance of our approach to group fragmented layers. The experimental results demonstrate that our approach achieves state-of-the-art performance. Additionally, a user study is conducted to prove that our approach can assist in generating high-quality front-end code under a real-world application scenario.

Our code and data are available at \url{https://github.com/zjl12138/ULDGNN}. The contributions of this study can be summarized as follows:
\begin{itemize}
    \item We propose a novel approach to tackle the fragmented layer grouping problem by classifying layers and detecting the bounding boxes of merging groups. In addition, post-process algorithms are developed to group semantically consistent fragmented layers inside each bounding box.
    \item We construct a graph representation for GUI design prototypes and adopt a graph neural network, which fully exploits the multimodal information, to learn a better representation vector for each layer.
    \item Experiments on a real-world dataset demonstrate our model achieves state-of-the-art performance. Additionally, an empirical study is conducted to validate that our approach can facilitate an intelligent downstream tool to generate more maintainable and readable front-end code. 
\end{itemize}

\section{literature review}\label{sec2}

\subsection{GUI Understanding}

GUI implementation is a time-consuming process, which prevents developers from devoting the majority of time to developing unique features of applications. It attracts researchers who adopt deep learning techniques to understand GUI and automate the developing process. Previous work can be summarized based on the resource where GUI originated from. Understanding GUI from screenshots or wireframe sketches is an active research field. For example, Magic layouts \cite{manandhar2021magic} detects and classifies UI components from UI images. Pix2code and doodle2app attempted to automate code generation from GUI wireframes or screenshots \cite{beltramelli2018pix2code,mohian2020doodle2app}. Previous methods already keep an eye on fully exploiting rich multimodal information to understand GUI \cite{he2021actionbert,liu2018learning,multimodal_icon_annotation}. Several papers or tools focus on improving GUI understanding and generation from original design prototypes created in design software (e.g. Sketch, Figma, PhotoShop, etc). These studies highly correlate with our work. UILM \cite{UILM} extracts the boundary information of UI layers from original design prototypes and composite them with the GUI screenshots to create a half-semantic image. With the boundary prior, UILM can regress more accurate bounding boxes for merging groups. EGFE \cite{egfe} flattens UI layers into a sequence and classifies each sequence element with a transformer. Imgcook \cite{imgcook} is a popular platform developed by the Alibaba Group to generate front-end code automatically from design files.

\subsection{GUI grouping}

Forming GUI groups of different granularities are used for intelligence to automate GUI testing \cite{degott2019learning, humanoid}, implementation, and automation tasks. To our knowledge, previous grouping methods can be summarized as three types: component-level, section-level, and layer-level. In the component-level category,  UIED \cite{UIED2} detects and forms perceptual groups of GUI widgets based on a psychologically-inspired, unsupervised visual inference method. Some methods \cite{beltramelli2018pix2code, GUI_skeleton} adopt image captioning models to generate GUI view hierarchy from GUI images. Methods in the section-level category group GUI elements into tab, bar, or layout sections. For example, REMAUI \cite{Revers_engi} group GUI widgets to three Android-specific layouts. Screen recognition \cite{screen_recog} develops some heuristics for inferring tab and bar sections. Xiao et al. proposed the semantic component group \cite{xiao2022ui, xiao2024ui} to identify UI components that achieve certain interaction functions or visual information. They introduced a vision detector based on deformable-DETR for semantic component grouping, aiming to improve the performance of multiple UI-related software tasks \cite{xiao2024ui}. While these methods are based on GUI design images, there are some GUI implementation-oriented methods to group GUI elements into sections. For example, ReDraw \cite{moran2018machine} and FaceOff \cite{faceoff} solve the layout problem by finding in the codebase the layouts containing similar GUI widgets. Some methods adopt specific layout algorithms to synthesize modular GUI code or layout \cite{gen_reu_webcomp, Robust_Relational_Layout} and infer GUI duplication \cite{Near-Duplicate}.

Recently, the fragmented layer grouping problem, which requires grouping low-granularity layers in GUI design prototypes during industrial GUI development, has attracted researchers' attention. Deep-learning-based techniques, such as object detection \cite{UILM} and transformer \cite{egfe}, are adopted to group fragmented layers to facilitate an intelligent downstream tool to generate more maintainable and readable code. However, these methods have some limitations and drawbacks as described in the introduction. In this study, we propose a new algorithm to tackle fragmented layer grouping, which outperforms previous work.

\subsection{Graph Neural Networks}

Recent years have witnessed a great surge of promising graph neural networks (GNNs) being developed for a variety of domains including chemistry, physics, social sciences, knowledge graphs, recommendations, and neuroscience. Graph learning refers to the process of learning from graph-structured data by leveraging the relationships between nodes and edges to capture both local and global patterns. The first GNN model was proposed in \cite{gori2005new}, which is a trainable recurrent message-passing process. To generalize the convolution operation to non-Euclidean graphs, these works \cite{bruna6203spectral,defferrard2016convolutionalneuralNetworkongraphs,kipf2022semi} defined spectral filters based on the graph Laplacian matrix. Spatial-based models define convolutions directly on the graph vertexes and their neighbors. Monti et al. \cite{monti2017geometric} presented a unified generalization of CNN architectures to graphs. Hamilton et al. \cite{GraphSAGE} introduced GraphSAGE, a method for computing node representations in an inductive manner that operates by sampling a fixed-size neighborhood of each node and performing a specific aggregator over it. Some methods attempted to enhance the original models with anisotropic operations on graphs, such as attention \cite{velivckovicgraph,brodyattentive} and gating mechanisms \cite{ruiz2020gated}. Xu et al. \cite{xu2018GIN} aimed at improving upon the theoretical limitations of the previous model. Li et al. \cite{li2019deepgcns} and Chen et al. \cite{chen2020GCNII} tried to overcome the over-smoothing problem when GCN goes deeper, and some current work already attempts to introduce transformer network into graph learning \cite{rampavsek2022recipe,ying2021transformers}.

Researchers also have great interest in utilizing graph neural networks to tackle computer vision tasks, such as 3D object detection \cite{shi2020pointGNN}, skeleton-based action recognition \cite{wen2019graphskeleton}, and semantic segmentation \cite{qi20173d}. Graph neural networks can also be helpful for GUI understanding. Ang et al. \cite{HAMP} combines
graph neural networks with scaled dot-product attention to learn the embeddings of heterogeneous nodes in GUI designs, which achieves state-of-the-art performance in UI representation learning tasks. Li et al. \cite{layout_dinoiser} introduce GNNs for multi-class node classification used for denoising GUI view hierarchy.
Inspired by these studies, we also attempt to introduce a graph neural model to our proposed pipeline to group fragmented layers in the UI design prototypes. More details are described in Section \ref{Approach}.

\section{Approach}\label{Approach}
\subsection{Pipeline Overview}
\begin{figure*}[htbp]
\centering 
\includegraphics[width=1\textwidth]{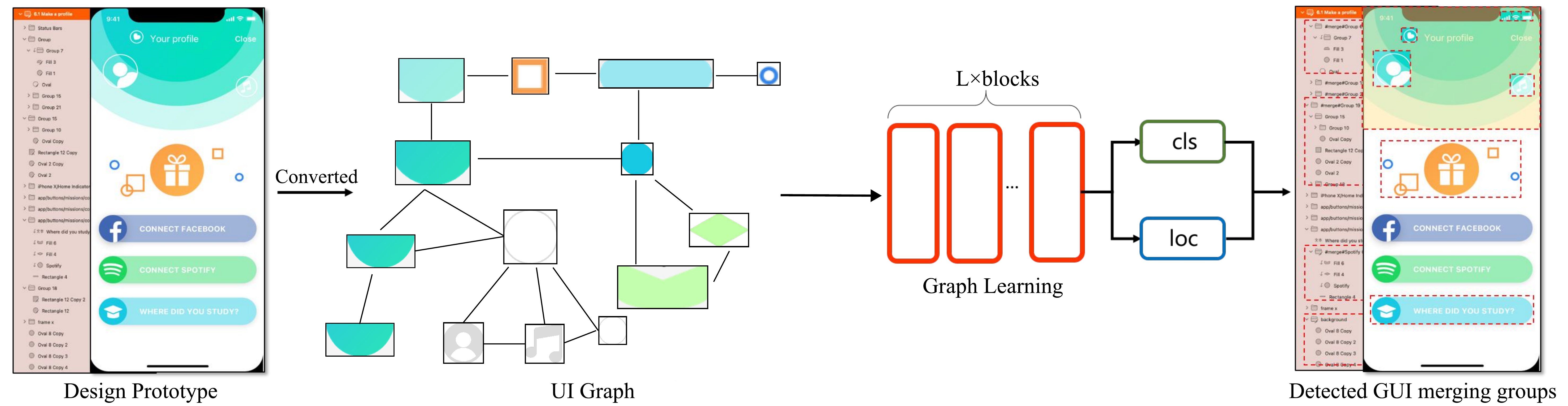} 
\caption{\textbf{Overview of the proposed method.} By accessing the view hierarchy of the design prototype, we construct a UI graph based on the geometric relationships between layers. We propose a graph learning block to capture the semantic associations and spatial structure between layers. We design a layer classification branch (cls) and a bounding box regression branch (loc), which are used to classify layers and regress the bounding boxes of merging groups, respectively.} 
\label{pipeline} 
\end{figure*}

\begin{figure*}[ht]
\centering 
\includegraphics[width=1\textwidth]{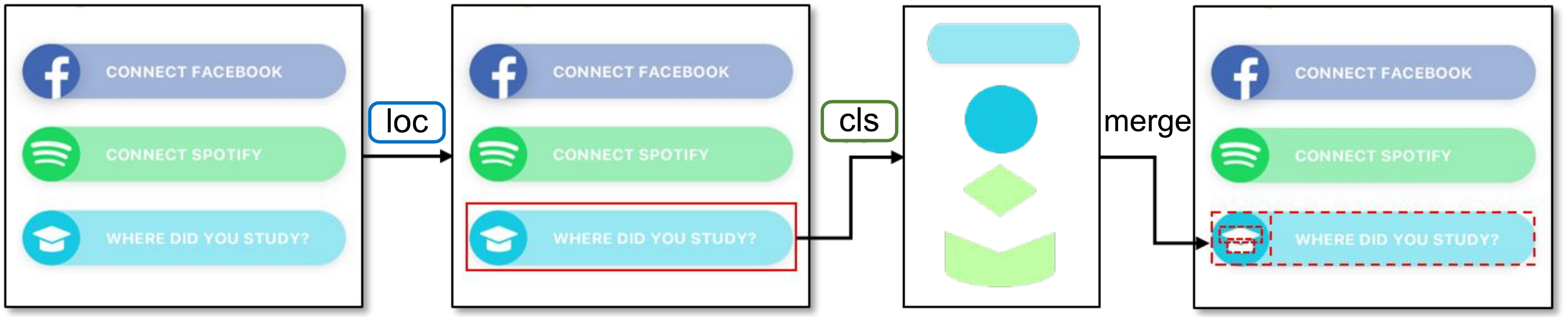} 
\caption{\textbf{The process of fragmented layer grouping.} To group fragmented layers, we first localize merging groups and classify layers to determine if they are fragmented layers. Then, within the bounding boxes of the detected merging groups, we group the fragmented layers that need to be merged. We use solid red lines to represent the predicted bounding boxes of merging groups and dashed red lines to represent the fragmented layers.} 
\label{frag_grouping} 
\end{figure*}

The pipeline of our approach is visualized in Figure \ref{pipeline} and Figure \ref{frag_grouping}. A design artboard is composed of UI layers that draw various UI components. We parse a design artboard into a layer list containing the following multimodal information, $\{(x, y, w, h), \textit{img-tensor}, \textit{category}\}^{n}_{i=1}$, in which $(x, y, w, h)$ is the layer’s wireframe information, img-tensor is the layer’s image with a resolution of 64$\times$64, category denotes the type of a layer (such as Text, ShapePath, etc). We encode all the attributes into embedding vectors and sum them up to obtain the initial representation vectors for each layer. Our graph-learning-based model refines and updates the layer's representation vectors. It predicts whether a layer is fragmented and regresses a bounding box for each fragmented layer. After a box merging algorithm, we obtain the final results of bounding boxes representing merging groups' boundaries. Fragmented layers inside the same bounding box are considered to be grouped. Specifically, our algorithm outputs a merging group list formulated as $[\{layer_{j}\}_{j \in g_{i}}]^{N}_{i=1}$, where \textit{N} denotes the number of merging groups, $g_{i}$ is the set of fragmented layer IDs in this merging group. Based on the output, our method can construct an optional relation matrix (used for evaluation) $M_{n\times n}$, where $M_{ij} = 1$ denotes $layer_{i}$ and $layer_{j}$ lie inside the same UI component, and $M_{ij} = 0$ separate layers with different semantics from each other.

\subsection{Graph Construction and Feature Extraction}

\subsubsection{Graph Construction}

In fragmented layer grouping, the core task of the proposed graph neural network is to discover the semantic consistency between fragmented layers. To more comprehensively explore the semantic similarity of GUI layers, we constructed a graph model based on the wireframe attributes of layers and enhanced the representation capabilities of node embeddings through the GNN. Our goal is to leverage the strengths of the GNN to facilitate information flow between layers within the same group, enabling a deeper understanding of the entire GUI layout. Through K iterations of the GNN, each GUI layer can aggregate features from its K-hop neighbors, helping it better understand its semantic role within the entire graph.

\begin{figure*}[htbp]
\centering 
\includegraphics[width=1\textwidth]{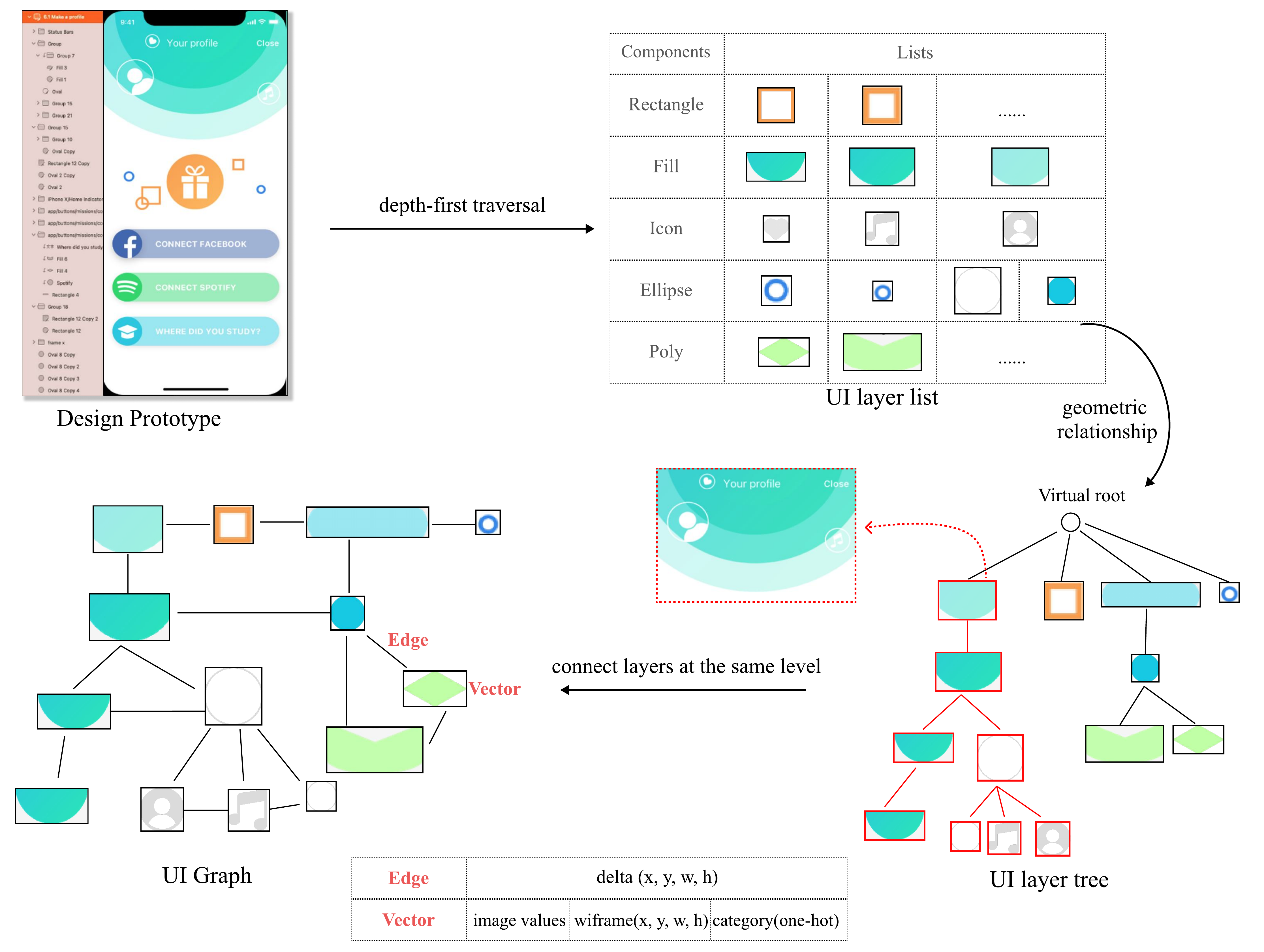} 
\caption{\textbf{The workflow of UI graph constructing.}}
\label{graph} 
\end{figure*}

As shown in Figure \ref{graph}, given a design prototype, we use depth-first traversal to obtain a \textit{flattened layer list}. Then, we sort the layer list in descending order based on the area of each layer and determine the parent-child relationships according to the inclusion relationships between layers, thus constructing a \textit{layer tree}. For example, when \textit{layer A} completely contains \textit{layer B and layer C}, A is considered the parent node of B and C, while B and C are sibling nodes. It is important to note that even if layer B further contains \textit{layer D}, D is a child of B, not A. This ensures clarity and logical consistency in the hierarchical structure. By leveraging the intrinsic relationships between layers, this layer tree eliminates manually introduced hierarchical errors in the design prototype. Next, we construct edges between all layers at the same level in the layer tree and remove the virtual root to obtain the \textit{UI graph}.

\subsubsection{Multimodal Attributes Encoding}

Unlike methods that solely parse GUI components from screenshots \cite{manandhar2021magic, multimodal_icon_annotation}, we leverage the rich multimodal information embedded within original design prototypes to address the fragmented layer grouping task. This subsection elaborates on how we encode multimodal attributes in detail.

Visual information plays a pivotal role in fragmented layer grouping. To balance time and space complexity, we employ a pre-trained ResNet-50 model as the backbone to encode $64 \times 64$ layer images into visual feature vectors. In GUI design software, each layer is categorized to help designers create geometric shapes (e.g., Oval, Rectangle) or other GUI elements (e.g., text and images). There are 13 layer categories, and we learn an embedding matrix $\mathcal{E}_{13\times d}$, where each row represents the embedding vector of a corresponding category. For wireframe information, we convert 4-dimensional coordinates into high-dimensional vectors using high-frequency functions as described in \cite{mildenhall2020nerf}:

\begin{equation} \gamma(x) = (\sin(2^{0}\pi x), \cos(2^{0}\pi x), \ldots, \sin(2^{L-1}\pi x), \cos(2^{L-1}\pi x)) \label{freq-func} \end{equation}

We then use a parameter matrix $\mathcal{M}_{8\times L, d}$ to embed the high-dimensional vector into a $d$-dimensional space. While several approaches aim to design more sophisticated multimodal feature fusion strategies, we employ a simple yet empirically effective strategy by directly adding these feature embedding vectors.

Additionally, we assign a feature vector to each edge. For an edge connecting nodes $v_{i}$ and $v_{j}$, we utilize the high-frequency function in formula \ref{freq-func} to encode the differences in wireframe coordinates, $(\Delta x, \Delta y, \Delta w, \Delta h)$, as the edge attribute vector. We acknowledge that layers within the same group tend to be spatially proximate. In fact, fragmented layer groups can be further divided into two categories: those where the internal layers are spatially adjacent (e.g., icon layers) and those where the internal layers are spatially distant (e.g., background layers). By encoding the spatial distances between layers and incorporating this information as edge embeddings, we significantly enhance the performance of the Graph Neural Network. This encoding enables the model to more effectively capture and differentiate the spatial relationships between layers, which is crucial for accurate layer grouping. We have integrated an attention mechanism within the GNN to focus on layer groups that may exhibit larger spatial separations but still demonstrate underlying correlations. This attention mechanism further improves the model's representational capacity and overall accuracy, ensuring that both spatially close and distant layer groups are appropriately identified and merged.

In summary, as depicted in Figure \ref{graph}.(d), each node in the graph is represented by a multimodal feature embedding, incorporating image features, wireframes, and category information. Each edge corresponds to a feature embedding that encodes the spatial relationships between adjacent nodes.

\subsection{Network Architecture and Loss Functions}
\subsubsection{Graph Learning Blocks}

\begin{figure*}[htbp]
\centering 
\includegraphics[width=0.3\textwidth]{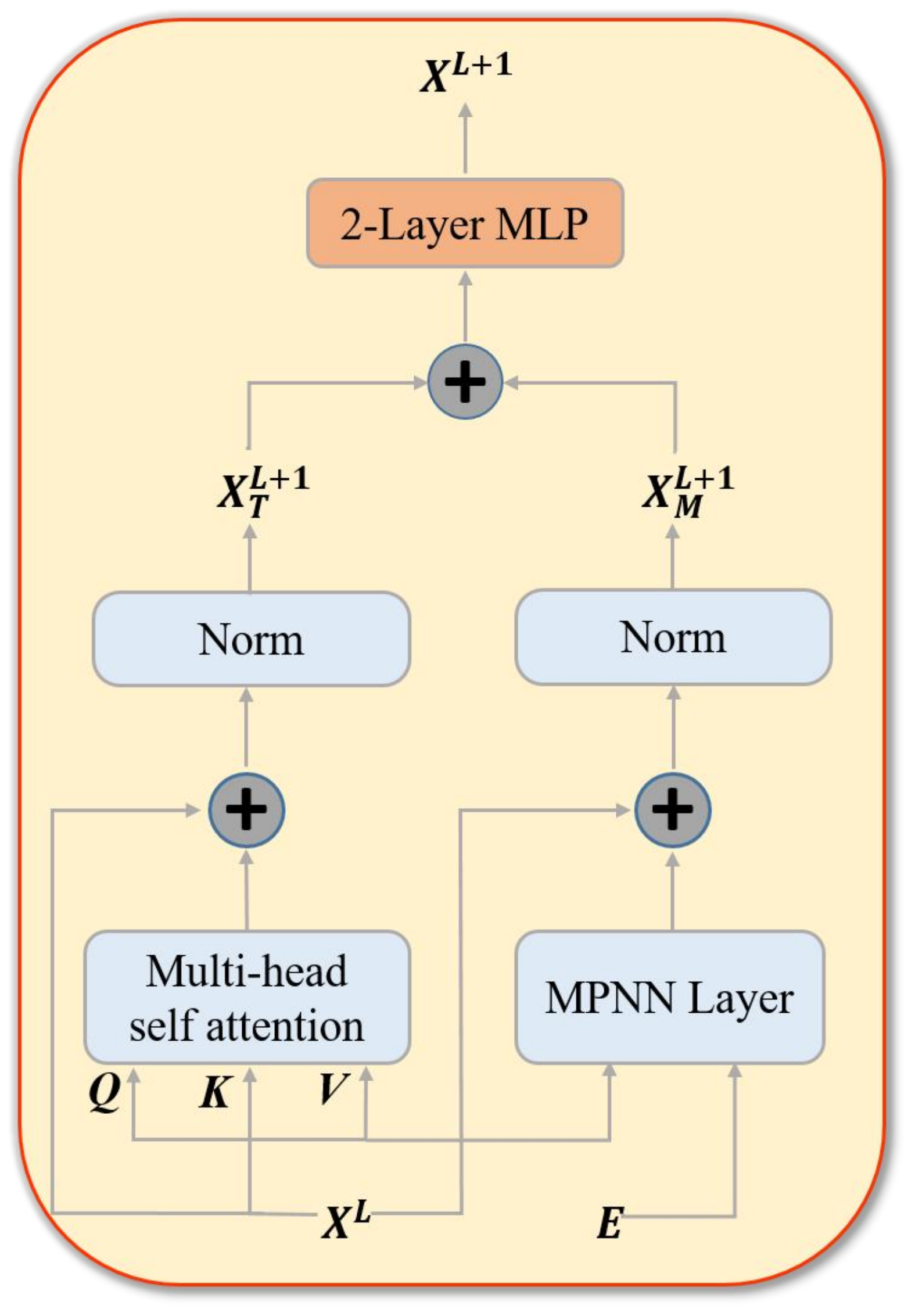} 
\caption{\textbf{Details of graph learning blocks:} Inspired by \cite{rampavsek2022recipe}, we introduce a multi-head attention module to our graph learning blocks to break through the fundamental limitations of GNNs.} 
\label{graphblock} 
\end{figure*}

As shown in Figure \ref{graphblock}, the proposed graph learning module consists of a multi-head self-attention module and a message-passing neural network (MPNN) layer. Graph neural networks (GNNs) can be formalized within the message-passing framework, where node representations are updated through the following iterative formula:

\begin{align}
h^{(k)}_{v} &= \text{COMBINE}^{(k)}\left( h^{(k-1)}_{v}, \right. \notag \\
&\quad \left. \text{AGGREGATE}^{(k)}\left( \left\{ \text{MESSAGE}^{(k)}\left( h^{(k-1)}_{v}, h^{(k-1)}_{u}, e_{uv} \right) \mid u \in \mathcal{N}_{v} \right\} \right) \right)
\end{align}

In the \(k\)-th iteration, for each node \(v\), the GNN first computes messages from its neighboring nodes \(u \in \mathcal{N}_{v}\) and the associated edge attributes \(e_{uv}\), and then aggregates these messages. The aggregated information is combined with the node’s previous representation \(h^{(k-1)}_{v}\) to update its representation \(h^{(k)}_{v}\).

After \(k\) iterations, the representation of node \(v\) captures the structural information within its \(k\)-hop neighborhood. To ensure consistency, the \(\text{AGGREGATE}\) function must be permutation-invariant to the order of neighboring nodes, while the \(\text{COMBINE}\) function should be differentiable to facilitate gradient-based optimization. In our implementation: The \(\text{MESSAGE}\) function is defined as:
    \begin{align}
        \text{MESSAGE}^{(k)}\left( h^{(k-1)}_{v}, h^{(k-1)}_{u}, e_{uv} \right) = h^{(k-1)}_{u}
    \end{align}
    
    which directly passes the features of the neighboring nodes. 
    
    The \(\text{AGGREGATE}\) function sums the messages:
    \begin{align}
            \text{AGGREGATE}^{(k)}\left( \left\{ h^{(k-1)}_{u} \mid u \in \mathcal{N}_{v} \right\} \right) = \sum_{u \in \mathcal{N}_{v}} h^{(k-1)}_{u}
    \end{align}
    The \(\text{COMBINE}\) function is implemented as a two-layer multilayer perceptron (MLP):
    \begin{align}
    h^{(k)}_{v} = \text{MLP}^{(k)}\left( h^{(k-1)}_{v} + \sum_{u \in \mathcal{N}_{v}} h^{(k-1)}_{u} \right)
    \end{align}
    This design follows the Graph Isomorphism Network (GIN) framework \cite{xu2018GIN}, which has expressive power equivalent to the Weisfeiler-Lehman (WL) graph isomorphism test.

However, traditional GNN methods face issues such as over-smoothing and over-compression \cite{chen2020GCNII}. To address these limitations, recent studies have introduced global attention mechanisms, allowing nodes to attend to all other nodes in the graph. Inspired by \cite{rampavsek2022recipe}, we incorporated multi-head self-attention into our model. The node representations are updated according to the following formulas:

\begin{align}
    \mathbf{X}_{\text{MPNN}}^{(L+1)} &= \text{MPNN}^{(L)}\left( \mathbf{X}^{(L)}, \mathbf{E}^{(L)} \right), \\
    \mathbf{X}_{\text{Attn}}^{(L+1)} &= \text{SelfAttn}^{(L)}\left( \mathbf{X}^{(L)} \right), \\
    \mathbf{X}_{\text{MPNN}}^{(L+1)} &= \text{LayerNorm}\left( \text{Dropout}\left( \mathbf{X}_{\text{MPNN}}^{(L+1)} \right) + \mathbf{X}^{(L)} \right), \\
    \mathbf{X}_{\text{Attn}}^{(L+1)} &= \text{LayerNorm}\left( \text{Dropout}\left( \mathbf{X}_{\text{Attn}}^{(L+1)} \right) + \mathbf{X}^{(L)} \right), \\
    \mathbf{X}^{(L+1)} &= \text{MLP}^{(L)}\left( \mathbf{X}_{\text{MPNN}}^{(L+1)} + \mathbf{X}_{\text{Attn}}^{(L+1)} \right),
\end{align}

Where \( \mathbf{X}^{(L)} \) represents the node feature matrix at the \(L\)-th layer. \( \mathbf{E}^{(L)} \) represents the edge attribute matrix at the \(L\)-th layer. \( \text{MPNN}^{(L)} \) captures local structural information through message passing. \( \text{SelfAttn}^{(L)} \) captures global context information through multi-head self-attention. \( \text{Dropout} \) and \( \text{LayerNorm} \) are used for regularization and training stability. \( \text{MLP}^{(L)} \) is a feedforward network that fuses information from both modules.

By integrating local message passing and global self-attention, our model captures comprehensive structural information, addressing the issues of over-smoothing and limited expressiveness in traditional GNNs. This hybrid approach leverages the strengths of both mechanisms, leading to improved performance in graph-based tasks.

\subsubsection{Classification and Boundary Regression}

After K iterations of updating node representation, the final node embedding vectors are fed into two MLP branches. The first branch predicts whether a layer is fragmented or not, and the second branch localizes the boundary of the merging groups. The classification branch roughly divides all GUI layers in the design artboard into two categories, which filter out the majority of unrelated layers for further grouping work. GUI layers that are filtered out may already contain rich semantics to be a GUI component. To group fragmented layers, we adopt the localization branch to first regress the bounding box of merging groups and then those layers inside it can naturally be grouped. More details about GUI layer grouping can be found in the next section.

The challenge for object detection is that the target object's size is not fixed and can vary greatly. In our fragmented layer grouping task, some UI merging groups draw large background patterns and some draw small icons. Inspired by Faster-RCNN \cite{girshick2015fast}, we also pre-define a set of anchor boxes with different aspect ratios and sizes to tackle the situation mentioned above. Different from object detection algorithms, which set the centers of anchor boxes to pixels, we directly regard the centers of UI layers to be the centers of anchor boxes. We choose three aspect ratios, which are 2:1, 1:1, and 1:2. The heights for each aspect ratio are 16, 128, and 256 respectively. The localization branch predicts the offsets of deforming these anchor boxes to the ground truth. We also predict a confidence score for each predicted bounding box and an NMS algorithm is utilized to obtain the final bounding box regression results. Before the NMS algorithm, we first filter out the bounding boxes that have non-maximum confidence scores for each fragmented layer. 

\subsubsection{Loss Functions}

The total loss function \(\mathcal{L}_{\text{total}}\) in our model is a weighted sum of three components: the classification loss \(\mathcal{L}_{\text{cls}}\), the localization loss \(\mathcal{L}_{\text{loc}}\), and the confidence loss \(\mathcal{L}_{\text{con}}\).

where \(\lambda_{\text{cls}}\), \(\lambda_{\text{loc}}\), and \(\lambda_{\text{con}}\) are hyperparameters that balance the contribution of each loss term.

To address the class imbalance between positive and negative samples, we employ the focal loss \cite{ross2017focal}. The focal loss modulates the standard cross-entropy loss by adding a factor that down-weights easy examples and focuses training on hard negatives. It is defined as:

\begin{equation}
    \mathcal{L}_{\text{cls}} = -\alpha_t (1 - p_t)^\gamma \log(p_t),
\end{equation}

where:

\begin{equation}
    p_t = \begin{cases}
    p, & \text{if } y = 1, \\
    1 - p, & \text{if } y = 0,
    \end{cases}
\end{equation}

and \(p \in [0,1]\) is the predicted probability for the class with ground-truth label \(y \in \{0,1\}\). \(\alpha_t \in [0,1]\) is a weighting factor for class \(t\) to address class imbalance. \(\gamma \geq 0\) is the focusing parameter that adjusts the rate at which easy examples are down-weighted.

By modulating the loss with \((1 - p_t)^\gamma\), the focal loss focuses learning on hard examples where \(p_t\) is small, thus improving the model's performance on imbalanced datasets.

For bounding box regression, we utilize the Complete Intersection over Union (CIoU) loss \cite{zheng2020distance}, which enhances the standard IoU loss by incorporating the distance between the centers of the predicted and ground-truth boxes as well as the aspect ratio consistency. The CIoU loss is formulated as:

\begin{equation}
    \mathcal{L}_{\text{loc}} = 1 - \text{IoU}(\mathbf{b}, \mathbf{b}^{\text{gt}}) + \frac{d^2(\mathbf{b}, \mathbf{b}^{\text{gt}})}{c^2} + \alpha v,
\end{equation}

where \(\text{IoU}(\mathbf{b}, \mathbf{b}^{\text{gt}})\) is the Intersection over Union between the predicted bounding box \(\mathbf{b}\) and the ground-truth box \(\mathbf{b}^{\text{gt}}\). \(d(\mathbf{b}, \mathbf{b}^{\text{gt}})\) is the Euclidean distance between the centers of \(\mathbf{b}\) and \(\mathbf{b}^{\text{gt}}\). \(c\) is the diagonal length of the smallest enclosing box that covers both \(\mathbf{b}\) and \(\mathbf{b}^{\text{gt}}\). \(v\) measures the discrepancy between the aspect ratios of the predicted and ground-truth boxes:

\begin{equation}
    v = \frac{4}{\pi^2} \left( \arctan\left( \frac{h^{\text{gt}}}{w^{\text{gt}}} \right) - \arctan\left( \frac{h}{w} \right) \right)^2,
\end{equation}

\begin{equation}
    \alpha = \frac{v}{(1 - \text{IoU}(\mathbf{b}, \mathbf{b}^{\text{gt}})) + v},
\end{equation}

where \(w\) and \(h\) are the width and height of the predicted box, and \(w^{\text{gt}}\) and \(h^{\text{gt}}\) are those of the ground-truth box.

The term \(\frac{d^2(\mathbf{b}, \mathbf{b}^{\text{gt}})}{c^2}\) penalizes the distance between the centers, encouraging better localization, while \(\alpha v\) adjusts for differences in aspect ratios, promoting boxes with similar shapes to the ground truth.

To supervise the confidence prediction for each bounding box, we align the predicted confidence score \(\hat{C}\) with the IoU between the predicted box and the ground-truth box. We employ the \(L_1\) loss:

\begin{equation}
    \mathcal{L}_{\text{con}} = \left| \hat{C} - \text{IoU}(\mathbf{b}, \mathbf{b}^{\text{gt}}) \right|.
\end{equation}

This loss encourages the model to produce confidence scores that are consistent with the actual localization accuracy, thereby improving the reliability of the confidence estimation.

By combining these components, the total loss function effectively balances classification accuracy, localization precision, and confidence estimation. The hyperparameters \(\lambda_{\text{cls}}\), \(\lambda_{\text{loc}}\), and \(\lambda_{\text{con}}\) are typically set based on validation performance and can be adjusted to prioritize different aspects of the model's performance.

\begin{equation}
    \mathcal{L}_{\text{total}} = \lambda_{\text{cls}} \mathcal{L}_{\text{cls}} + \lambda_{\text{loc}} \mathcal{L}_{\text{loc}} + \lambda_{\text{con}} \mathcal{L}_{\text{con}}.
\end{equation}

\subsection{Box merging and UI Layer Grouping}

\subsubsection{NMS Algorithm}
Our model regresses a bounding box for each fragmented layer. Many boxes represent the same merging area boundary. We design an algorithm to obtain a final bounding box based on a non-maximum-suppress (NMS) algorithm that is used in the object detection task. The motivation for our algorithm has two aspects, the first one is that our final result box should be large enough so that it can contain the whole UI components. It doesn’t matter that our predicted boxes’ are not so accurate. On the other side, simply discarding boxes that don’t have maximum confidence scores may result in removing accurate boxes that have a comparable confidence score with the remaining ones.

For the motivation above, we propose an algorithm calculating a final merged box by considering the entire overlapped box cluster, as described in Algorithm.\ref{alg1}. We first sort the bounding box in descending order according to the confidence scores. For the box with the highest confidence score, we calculate the IoU of this box and other boxes and find those boxes whose IoU values are larger than a pre-defined threshold value. For this overlapped box cluster, we use the average size of these boxes to be the final result of this corresponding UI group’s boundary.

\subsubsection{Fragmented Layer Grouping}

It is trivial to group fragmented layers because our model regresses the bounding boxes of merging groups. Considering the overlap between background UI components and others, we sort merged boxes in ascending order according to the rectangle area and group semantically consistent layers inside smaller boxes first. We traverse the sorted bounding box list and calculate the proportion of fragmented layers that intersect with the current box. If the value exceeds the threshold, we assign this fragmented layer to the current merging groups. In this way, we can effectively group fragmented layers in a design prototype and facilitate a downstream code-generation platform to generate code of higher quality. There is another advantage of our algorithm, which is the capability of correcting errors in grouping results based on some prior knowledge. We observe that a Text layer is usually not grouped with others, while layers depicting geometric shapes are part of merging groups. When searching inside merged boxes, we can consider grouping layers that are predicted as non-fragmented but belong to a specific category (such as Oval, Rectangle, etc). 
\begin{algorithm}[H]
\caption{Our improved NMS algorithm}\label{alg:alg1}
\begin{algorithmic}
\State 
\State {\textsc{\textbf{Input}}} $\mathcal{B}=\{\textbf{b}_{1},...,\textbf{b}_{n}\}, \mathcal{D}=\{d_{1},...,d_{n}\}$
\State $\mathcal{B}$ is the set of detected bounding boxes
\State $\mathcal{D}$ is the confidence scores of corresponding boxes
\State $\mathcal{M} \gets \{ \}$, $\mathcal{S} \gets \{ \}$
\While{ $\mathcal{B} \neq \emptyset $ }
 \State $i \gets argmax(\mathcal{D})$
 \State $\mathcal{C}\gets \emptyset$
 \For{$b_{j} \in \mathcal{B}$}
 \If{IoU($b_{j}, b_{i}$) $ > thr$}
   \State $\mathcal{C} \gets \mathcal{C} \cup \{b_{j}\}$
   \State $\mathcal{B} \gets \mathcal{B}-\{b_{j}\}$, $\mathcal{D} \gets \mathcal{D}-\{d_{j}\}$
 \EndIf
 \EndFor
 \State $\mathcal{M} \gets \mathcal{M} \cup average(\mathcal{C})$
 \State $\mathcal{S} \gets \mathcal{S} \cup average(\{d_{k}|b_{k} \in \mathcal{C}\})$
 \EndWhile       

\State \textbf{return}  $\mathcal{M}, \mathcal{S}$
\end{algorithmic}
\label{alg1}
\end{algorithm}

\section{Experiments}
\subsection{Implementation Details}
\subsubsection{Data preparation}
Following UILM \cite{UILM} and EGFE \cite{egfe}, we conduct experiments on their collected real-world design artboards to validate our algorithm’s effectiveness. The number of fragmented layers varies greatly in different design artboards. Therefore, we sort the data according to the number of fragmented layers, after which we split the dataset into three even collections. For each collection, we split it into a training set and a test set in a consistent proportion of 8:2. Considering that a design artboard may contain a substantial amount of layers, which can result in consuming too many computing resources during graph learning, we use a sliding window with a fixed window size to cut the artboard. Graphs are constructed for layers inside each window. 
\subsubsection{Training Details}
The multimodal attributes of layers are embedded into 128-dim feature vectors. We adopt ResNet50 \cite{he2016deep} pre-trained on ImageNet to extract the layer’s visual image features and only train a linear layer to transform the features into a 128-dim vector. For the high-frequency encoding, we set L to be 9 and utilize a linear layer to encode the result into an initial vector. The attention module of each layer has 4 attention heads and a hidden dimension of 128. We adopt the GINE \cite{hustrategies} layer as the graph neural module to learn neighborhood information around each node, and the hidden dimension is 128 as well. We conduct hyperparameter experiments to determine the optimal number of graph learning blocks. As shown in Table \ref{tab:ablation_hyperparm}, our model achieves the best performance when the number of layers is set to 9. The classification branch and the localization branch both consist of a 3-layer MLP, where the hidden dimension is set to 256. For the loss function, we set $\lambda_{cls}$ as 1, $\lambda_{loc}$ as 10.0, and $\lambda_{con}$ as 5.0 also based on the results of hyperparameter experiments (Table \ref{tab:ablation_hyperparm}). We set the IoU threshold for the NMS-fine algorithm to 0.45 and set the IoU threshold for merging associative layers to 0.7.

The whole algorithm is implemented by Pytorch and Pytorch-geometric library. All experiments are conducted on a Linux server with 4 Geforce-3090 GPUs. We train the cls branch for the first 50 epochs with AdamW optimizer \cite{AdamW} and the learning rate is set as 1e-4. Next, we train the two branches together for another 1000 epochs to converge. The initial learning rate is 1e-5, and it drops down exponentially every 10 epochs by 0.99. The training process takes around 20 hours. 

\subsection{Experimental Setting}

\subsubsection{Baselien Models}
UILM \cite{UILM} develops an objection detection method to detect the bounding boxes of merging groups based on the layer's boundary prior. Then all layers inside the same bounding box are grouped. 

EGFE \cite{egfe} adopts a transformer encoder to predict the label of each layer based on the multi-modal information. Then layers between two elements labeled ‘Start-merge’ in the sequence are considered to be merged into a UI merging group.

We also conduct experiments with previous graph neural networks. GCNII \cite{chen2020GCNII} is a state-of-the-art method inspired by ResNet. It introduces two key strategies: initial residual connection and identity mapping, which improve the shallow nature of GCN models and mitigate the over-smoothing problem. Similarly, GraphGPS \cite{rampavsek2022recipe} is a representative graph transformer framework that integrates positional/structural encoding, local message-passing, and global attention mechanisms. It achieves superior performance in graph learning tasks by effectively capturing both global and local structural features. In addition, we also attempt to replace the graph neural module in our method to validate its robustness. GCN \cite{kipf2022semi} and GAT \cite{brodyattentive} are utilized as two popular baseline models.

\subsubsection{Metrics}
Common metrics, which are precision, recall, f1-score, and accuracy, are used to evaluate the classification results. UILM cannot classify layers, so we considered a layer to be fragmented if 70\% of it lies inside the predicted bounding box. EGFE categorizes layers into three classes. Layers labeled 'Start-merge' and 'Merge' are considered to be fragmented because EGFE only groups layers of these two classes. \par
To evaluate the performance of the fragmented layer grouping task, we propose the following evaluation methods. The first one is that we evaluate the accuracy of predicting a similarity matrix, in which $M_{ij}$=1 denotes layer \textit{i} and layer \textit{j} belong to the same UI component, and $M_{ij}$=0 means layer \textit{i} and layer \textit{j} should be separated into different merging groups. We compare the predicted matrix with the ground-truth matrix to evaluate the layer grouping results of different methods. Specifically, the following formulas 
\begin{align}    
\text{\textit{asso-precision}} &= \frac{\underset{i\in S_{gt}}{\sum}\underset{j\in S_{gt}}{\sum}\mathbf{1}(\textbf{M}_{ij}==\textbf{M}^{gt}_{ij})}{(\#S_{gt})^2}) \label{asso-prec}\\
\text{\textit{asso-recall}} &= \frac{\underset{i\in S_{pred}}{\sum}\underset{j\in S_{pred}}{\sum}\mathbf{1}(\textbf{M}_{ij}==\textbf{M}^{gt}_{ij})}{(\#S_{pred})^2} \label{asso-recall}
\end{align}
evaluate the performance of our approach in two aspects, which is similar to precision and recall. Eq.\ref{asso-prec} shows how many correct merging pairs are predicted by the algorithms, and Eq.\ref{asso-recall} evaluates the quality of fragmented layer grouping results. In the formulas, $S_{pred}, S_{gt}$ denotes the set of layers in predicted and ground-truth merging groups respectively. $\#$\ denotes the number of a set. A design artboard may contain several merging groups, so we just average the results to obtain a final score. \par
Another evaluation method is calculating the IoU of predicted and ground-truth merging groups, and the formulas are,
\begin{align}
\text{\textit{iou-precison}} &= \underset{S_{pred}}{Average}(\max_{S_{gt}}\frac{\#(S_{pred}\cap S_{gt})}{\#(S_{pred}\cup S_{gt})})\\
\text{\textit{iou-recall}} &= \underset{S_{gt}}{Average}(\max_{S_{pred}}\frac{\#(S_{pred}\cap S_{gt})}{\#(S_{pred}\cup S_{gt})})
\end{align}
where $\#$ denotes the number of elements in the merging groups. For each predicted merging group, we calculate the IoU of every ground-truth merging group and itself, and we use the maximum value of these results to evaluate this merging group.

\subsection{Quantitative and Qualitative Analysis} 

\subsubsection{Comparison with Previous Methods} 

UILM adopts the 2D object detection model to locate the boundary of merging groups and find fragmented layers inside the area. However, when the bounding box of the merging groups is large, it will also contain other unrelated GUI layers that should not be merged into that merging group. Conversely, EGFE and our model classify each UI layer directly according to the multimodal information in the original design prototypes. Therefore, the UI layer classification accuracy of UILM is smaller than EGFE and our model, as shown in Table \ref{layer_classification}, due to retrieving many non-fragmented layers inside detected bounding boxes. The precision of EGFE outperforms our model a little, however, our model improves the recall by around 5.3\%. Furthermore, among graph-based methods, our approach also demonstrates the best performance, with an F1 score 2.2\% higher than GCNII and 1\% higher than GraphGPS. The primary issue with these methods is their over-reliance on the proper tuning of hyperparameters, which limits their ability to generalize to other graph tasks. These advanced mechanisms significantly amplify the influence of relationships between components on the layer classification results. Consequently, the features of one layer may excessively affect the connected or indirectly connected layers. Compared with previous methods such as EGFE and UILM, Table \ref{layer_classification} can validate the effectiveness of introducing graph neural modules to learn the relationship between UI layers. For example, the two graph neural models (GCN and GAT) can improve the recall and consequently, the f1-score of UI layer classification.  Our method also outperforms GCN and GAT, achieving F1 score improvements of 1.03\% and 1.01\%, respectively. Considering the performance of fragmented layer grouping, our method outperforms previous methods in all metrics greatly. We improve asso-precision and IoU-precision by 9.7\% and 6.7\% respectively compared with UILM. For asso-recall and IoU-recall, our method outperforms UILM by 3.2\% and 3.9\% respectively.

Here we further analyze the reported results above. UILM can detect accurate bounding boxes of merging groups but it may make mistakes for background layer grouping due to the visual overlap of layers. It assumes that all layers inside a bounding box belong to the corresponding merging group, which is the main reason for the low precision of grouping results. The main obstacle of EGFE is the class imbalance problem. It categorizes UI layers into 3 classes, which are 'Start-merge', 'merge', and 'None-merge'. However, the amount of 'Start-merge' equals the number of UI merging groups in the dataset, which is far fewer than the amount of the other two classes. The low merging precision of EGFE is attributed to the wrong prediction of the category 'Start-merge'.

Based on the analysis above, our algorithm not only inherits the advantages of EGFE and UILM but also avoids the weakness of these methods to enhance the whole performance on the fragmented layer grouping task.

\begin{table}[ht]
    \centering
    \caption{Classification results on the real-world dataset. We compare our algorithm with previous methods and other baseline modes.}
    \renewcommand\arraystretch{1.1}
    \begin{tabular}{l|c|c|c|c}
    \toprule
       \textbf{ Method }& \textbf{Precision} & \textbf{Recall} & \textbf{F1-score} & \textbf{Accuracy} \\
    \hline\hline
         UILM  & 0.741 & 0.844 & 0.776 & 0.877 \\
         EGFE  &\textbf{0.930} & 0.838 & 0.882 & 0.952 \\
         Attn+GCN & 0.900 & 0.880 & 0.890& 0.951 \\
         Attn+GAT & 0.906& 0.879& 0.892&0.953 \\
         GCNII & 0.895& 0.869& 0.881&0.942 \\
         GraphGPS & 0.911& 0.876& 0.893&0.954 \\
         Ours & 0.916 &\textbf{0.891} &\textbf{0.903} &\textbf{0.957} \\
    \bottomrule
    \end{tabular}
    \label{layer_classification}
    
\end{table}

\begin{table}[ht]
    \centering
    \caption{Evaluation on fragmented layer grouping. We report the four newly proposed metrics to compare our algorithm with other methods.}
    \renewcommand\arraystretch{1.1}
    \begin{tabular}{l|c|c|c|c}
    \toprule
       \textbf{ Method }& \textbf{Asso-prec.} & \textbf{Asso-rec.} & \textbf{IoU-prec.} & \textbf{IoU-rec.} \\
    \hline\hline
         UILM  & 0.721 &0.863& 0.697 & 0.704 \\
         EGFE  & 0.706 & 0.776 & 0.634 & 0.601  \\
         Attn+GAT &0.817 & 0.870& 0.759&  0.731\\
         Attn+GCN &0.811 & 0.864 &0.758 &0.727 \\
         Ours & \textbf{0.818}& \textbf{0.895}&\textbf{0.764} &\textbf{0.743} \\
    \bottomrule
    \end{tabular}
    \begin{tablenotes}
     \item[1] Asso-prec., Asso-rec., IoU-prec. and Iou-rec. are short for asso-precision, asso-recall, iou-precision, iou-recall respectively
    \end{tablenotes}
    \label{tab:eval_merging}
\end{table}

\subsubsection{Qualitative Analysis}
\begin{figure*}[htbp]
\centering 
\includegraphics[width=1\textwidth]{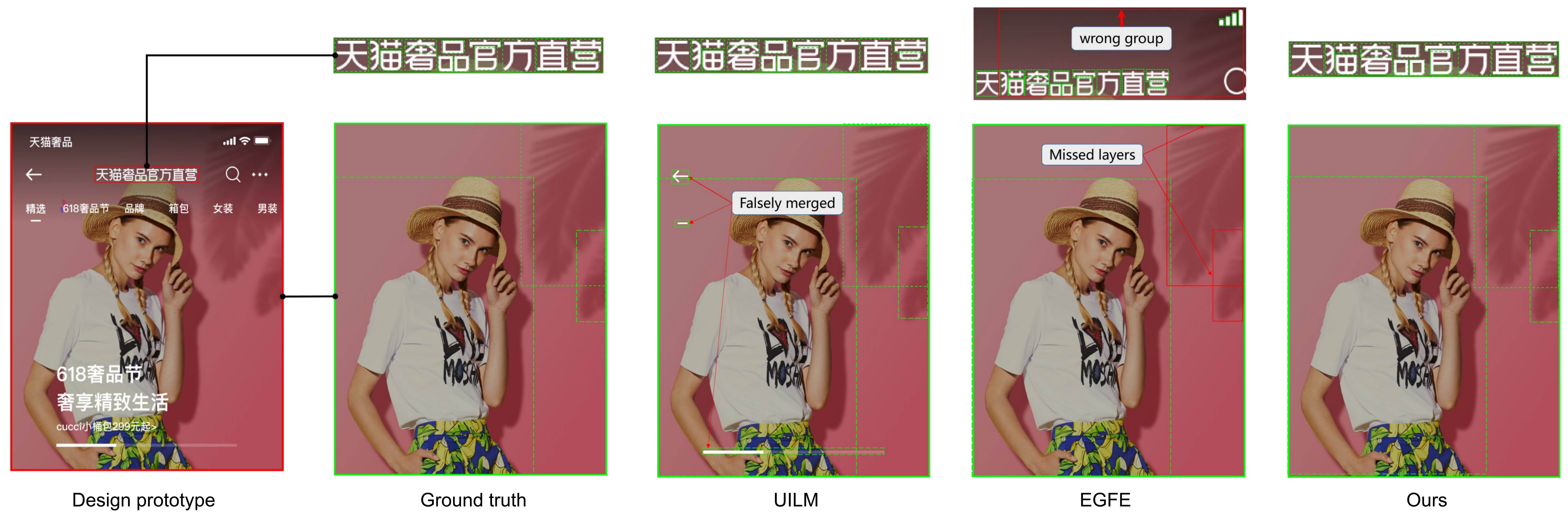} 
\caption{\textbf{Typical qualitative results:} In the design prototype, there are two typical merging groups highlighted by solid red lines: the decorative text component and the background component. UILM incorrectly merged three foreground components with the background component. Additionally, EGFE missed the layer of the background component and incorrectly identified the bounding box of the merging group (marked by the red box). Our method accurately identified both the decorative text component and the background component, demonstrating the effectiveness of our approach.} 
\label{qualitative} 
\end{figure*}

As shown in Figure \ref{qualitative}, we visualize a typical case to elaborate on the advantages of our algorithm over previous methods. UILM does well in detecting the bounding boxes of foreground merging groups. However, it is hard for UILM to distinguish between background and foreground layers directly due to the visual overlap. EGFE and our method can achieve higher accuracy in layer classification than UILM because we directly use the multi-modal information to classify layers. However, EGFE may sometimes miss some fragmented layers. For example, EGFE can not find two background layers (which depict the shadow of a tree). EGFE also makes mistakes in grouping stroke layers and signal layers due to the wrong prediction of class 'Start-merge'. To overcome the limitations of EGFE, we only categorize layers into two classes and further regress the boundary of merging groups. Quantitative and qualitative results demonstrate the effectiveness of grouping predicted fragmented layers inside predicted bounding boxes of merging groups.
Above all, our algorithm not only inherits the advantage of UILM to obtain a higher quality of grouping fragmented layers but also avoids the sample imbalance and low grouping recall of EGFE.

\subsection{Additional Study}
In this section, we present additional experimental results. We first discuss the effectiveness of our proposed strategies, including the layer graph construction based on wireframe coordinates, the self-attention module introduced in the graph learning blocks, the wireframe coordinates used as edge attributes, the improved NMS algorithm, which takes the entire overlapped box cluster into account, and the multimodal information used as input graph representation learning. To further validate the performance of our method on a wider range of real-world UI designs, we obtain an additional 508 design prototypes from the Figma community. Compared to our training dataset, these prototypes exhibit more variability in design quality and cover a broader range of business domains.

\subsubsection{Effectiveness of Layer Graph Construction}

In our approach, we start by reconstructing the structure of design prototypes based on the coordinates between layers. The idea is to break down all the original containers, allowing layers that are close in distance and spatially nested to aggregate in the graph, facilitating information flow between them during the learning process. At the same time, this approach helps eliminate irregularities in these designs, which often arise from designers' inconsistent layer organization or improper grouping of elements. To demonstrate its effectiveness, we compare it with the original view hierarchy. As shown in Table \ref{tab:ablation_1}, using the original structure led to a decrease in asso-recall and IoU-recall by approximately 8\% and 5\%, respectively.

\subsubsection{Effectiveness of Self-attention}

We remove the self-attention module to see how long-range dependency boosts the performance of our model. Due to the over-smoothing problem, we ultimately utilize a 5-layer GINE model for this experiment. As shown in Table \ref{tab:ablation_1}, the recall of fragmented layer prediction falls down by around 5\% while the recall of grouping UI layers also declines a lot. Inspired by \cite{rampavsek2022recipe}, global self-attention can help alleviate the over-smoothing, over-squashing, and other fundamental problems in graph neural networks. It is a bottleneck for GNNs to capture long-range dependency between nodes that are distant from each other in the graph. Self-attention mechanism requires nodes to attend the key-query process of all other nodes, which naturally addresses the issue of information communication bottleneck due to the limitation of graph topology. Overall, self-attention can not only improve the accuracy of retrieving fragmented layers but can improve the quality of UI layer grouping as well.  

\begin{table*}[htbp]
    \centering
    \caption{Additional Study on the design choices of our algorithm. We first investigate how edge attributes and self-attention boost performance. Then we evaluate the NMS algorithm more deeply.}
    \resizebox{\textwidth}{!} {
    \begin{tabular}{l|c|c|c|c|c|c|c|c} 
    \toprule
       \multirow{2}{*}{\textbf{ Method}} & \multicolumn{4}{c|}{\textbf{Classification}} & \multicolumn{4}{c}{\textbf{Grouping}} \\ \cmidrule{2-9}
       &\textbf{Precision} & \textbf{Recall} & \textbf{F1-score} & \textbf{Accuracy}  & \textbf{Asso-prec.} & \textbf{Asso-rec.} & \textbf{IoU-prec.} & \textbf{IoU-rec.} \\
    \midrule
    Edge with WH attr & 0.923 & 0.872 & 0.897 & 0.951 & 0.803 & 0.845 & 0.747 & 0.726 \\
    Edge with XY attr & 0.911 & 0.883 & 0.897 & 0.954 & 0.792 & 0.853 & 0.741 & 0.735 \\
    w/o Edge attr & \textbf{0.925} & 0.864 & 0.893 & 0.954 & 0.792 & 0.819 & 0.738 & 0.692 \\
    \midrule
    Original view hierarchy & 0.893 & 0.862 & 0.877 & 0.944 & 0.788 & 0.814 & 0.733 & 0.689 \\
    \midrule
     w/o Self-attention  & 0.886&0.843 &0.864 &0.941  &0.791 &0.841 & 0.721 &0.690\\
     Standard NMS & -&- &- &-  & 0.812& 0.871&0.748 &0.721\\
     w/o Box scoring & -& -&- &- &0.811 &0.868 & 0.738 &0.717\\
        Ours & 0.916 &\textbf{0.891} &\textbf{0.903} &\textbf{0.957}& \textbf{0.818}& \textbf{0.895}&\textbf{0.764} &\textbf{0.743}\\
    \bottomrule
    \end{tabular}
    }
    \label{tab:ablation_1}
\end{table*}
\subsubsection{Effectiveness of Edge Attributes}
To validate the effectiveness of edge attributes, we set up three baseline groups: edges with XY as attributes, edges with WH as attributes, and no encoding for edges. We believe that encoding coordinate information between UI layers is crucial because it signifies the type of merging area in UI layers. It is trivial that UI layers in UI icons lie very close while background layers spread around. Encoding the difference of adjacent nodes’ wireframe information as the edge attributes can facilitate our model to regress more accurate bounding boxes of merging groups. Table \ref{tab:ablation_1} shows that the quality of layer grouping can improve a lot after we encode the edge attributes, which proves the necessity of edge encoding during the graph learning process. Specifically, no encoding, only XY, and only WH as attributes result in a 7.6\%, 4.2\%, and 5.0\% decrease in asso-recall, and a 5.1\%, 0.8\%, and 1.7\% decrease in IoU-recall, respectively.

\subsubsection{Evaluation on Improved NMS Algorithm}
The third row of Table \ref{tab:ablation_1} shows the results of using a standard NMS algorithm to group fragmented UI layers. As described in Section \ref{Approach}, the key improvement is that we evaluate the whole overlapped box cluster to obtain the final box result. Instead, the original NMS algorithm discards overlapped boxes that have non-maximum classification probability or confidence score. The experimental results show that our algorithm can improve the asso-recall and IoU-recall f1-score by 2.4\% and 2.2\% respectively. It seems that the improvement is not very significant. One possible explanation is that we use the confidence score instead of the classification probability as the criteria to reserve the bounding box with the maximum value. The confidence score depicts the quality of the predicted bounding box more accurately than adopting the classification score. So the bounding box with the maximum confidence score is already accurate enough. The fourth row of Table \ref{tab:ablation_1} further reports the results of the original NMS algorithm that utilizes classification probability instead of predicted confidence score to non-maximum suppress overlapped boxes. The asso-recall and IoU-recall fall down by 2.7\% and 2.6\% respectively.  

\subsubsection{Effectiveness of Multimodal Information}

Table \ref{ablation_study_2} presents the experimental results of our model when visual features, categories, and wireframe information are removed from the input. The performance of our model decreases as multimodal information is removed, with the accuracy of UI layer classification remaining relatively stable while the performance of fragmented layer grouping is impaired. This may be due to inaccurate regressed bounding boxes of merging groups, which affects the quality of fragmented layer grouping results. 

Visual features have the greatest influence on the performance of our model among the three types of multimodal information, as observed from Table \ref{ablation_study_2}. Without integrating visual information, the classification accuracy of our model slightly degrades, but the asso-recall and IoU-recall decrease by 4.9\% and 5.1\%, respectively. This result is uniform with human perception as we predominantly understand UI based on screenshots rather than other underlying information. The category information about UI layers has less impact on our model due to the limited number of layer categories (only 13). Wireframe information plays a critical role in edge learning during the message-passing process, explaining the considerable performance drop across all metrics upon its removal.

\begin{table*}[htbp]
    \centering
    \caption{Ablation Study on multi-modal information. We remove the category, wireframe, and visual information consequently from the input to investigate how effective the multi-modal information is}
    \renewcommand\arraystretch{1.3}
    \scalebox{0.68}[0.68]{
    \begin{tabular}{l|c|c|c|c|c|c|c|c} 
    \toprule
       \multirow{2}{*}{\textbf{ Method}} & \multicolumn{4}{c|}{\textbf{Classification}} & \multicolumn{4}{c}{\textbf{Grouping}} \\ \cmidrule{2-9}
       &\textbf{Precision} & \textbf{Recall} & \textbf{F1-score} & \textbf{Accuracy}  & \textbf{Asso-prec.} & \textbf{Asso-rec.} & \textbf{IoU-prec.} & \textbf{IoU-rec.} \\
    \midrule
         w/o category  &0.904  & 0.878& 0.890&0.953 &0.800 &0.874 &0.747 &0.722\\
         w/o wireframe &0.916 &0.869 &0.890 &0.952 &0.790 &0.847 &0.741 &0.701 \\
         w/o image &0.883 & 0.870&0.876 &0.945 & 0.784& 0.846&0.725 &0.692\\
         Ours & \textbf{0.916} &\textbf{0.891} &\textbf{0.903} &\textbf{0.957}& \textbf{0.818}& \textbf{0.895}&\textbf{0.764} &\textbf{0.743} \\
    \bottomrule
    \end{tabular}
    }
    \label{ablation_study_2}
\end{table*}

\begin{table*}[htbp]
    \centering
    \caption{Additional Study on the hyper-parameters.}
    \resizebox{\textwidth}{!} {
    \begin{tabular}{l|c|c|c|c|c|c|c|c} 
    \toprule
       \multirow{2}{*}{\textbf{ Method}} & \multicolumn{4}{c|}{\textbf{Classification}} & \multicolumn{4}{c}{\textbf{Grouping}} \\ \cmidrule{2-9}
       &\textbf{Precision} & \textbf{Recall} & \textbf{F1-score} & \textbf{Accuracy}  & \textbf{Asso-prec.} & \textbf{Asso-rec.} & \textbf{IoU-prec.} & \textbf{IoU-rec.} \\
    \midrule
    1-layer & 0.873 & 0.858 & 0.865 & 0.926 & 0.791 & 0.862 & 0.739 & 0.715 \\
    3-layer & 0.881 & 0.862 & 0.871 & 0.933 & 0.809 & 0.875 & 0.746 & 0.721 \\
    6-layer & 0.902 & 0.889 & 0.895 & 0.946 & 0.812 & 0.889 & 0.751 & 0.729 \\
    12-layer & 0.911 & 0.890 & 0.900 & 0.953 & 0.816 & 0.892 & 0.759 & 0.742 \\
    \midrule
    Warm-up lr & 0.912 & 0.891 & 0.901 & 0.954 & 0.811 & 0.891 & 0.758 & 0.741 \\
    \midrule
    $\lambda_{\text{cls}} : \lambda_{\text{loc}} : \lambda_{\text{con}} = 1 : 1 : 1$ 
    & 0.905 & 0.881 & 0.893 & 0.953 & 0.803 & 0.875 & 0.749 & 0.728 \\
    $\lambda_{\text{cls}} : \lambda_{\text{loc}} : \lambda_{\text{con}} = 1 : 5 : 5$ 
    & 0.899 & 0.891 & 0.895 & 0.954 & 0.803 & 0.879 & 0.751 & 0.729 \\
    $\lambda_{\text{cls}} : \lambda_{\text{loc}} : \lambda_{\text{con}} = 1 : 5 : 10$ 
    & 0.901& 0.884 & 0.892 & 0.952 & 0.802 & 0.878 & 0.747 & 0.725 \\

    \midrule
    Ours & \textbf{0.916} &\textbf{0.891} &\textbf{0.903} &\textbf{0.957}& \textbf{0.818}& \textbf{0.895}&\textbf{0.764} &\textbf{0.743}\\
    \bottomrule
    \end{tabular}
    }
    \label{tab:ablation_hyperparm}
\end{table*}

\subsubsection{Evaluation on Figma Dataset}

To further validate the effectiveness of our method on a broader set of UI data, we collected additional design prototypes from the Figma community. Compared to the training data, these prototypes span a wider range of application scenarios, including shopping, finance, healthcare, and travel. Given the varying levels of expertise among community designers, these prototypes exhibit greater variability in quality and more pronounced structural errors compared to data from EGFE and UILM, which are derived from commercial applications developed by large enterprises. A total of 508 design prototypes were collected and subsequently inspected and annotated by five designers to label fragmented elements as ground truth. Our Figma dataset is comparable to the original test set, with one-hot encoding updated based on the layer categories in the Figma designs. These prototypes were used exclusively for testing to evaluate our model's generalization performance beyond the original training data. As shown in Table \ref{tab:ablation_figma}, our method maintained strong performance on the Figma dataset, with only a 4.0\% reduction in asso-recall and a 3.6\% reduction in IoU-recall compared to the original dataset. The performance decline is likely due to the significant distribution shift across different application scenarios in the UI design domain. Future research could explore domain adaptation techniques to further improve the model's robustness.

\begin{table*}[htbp]
    \centering
    \caption{Additional Study on the real dataset collected from Figma design.}
    \resizebox{\textwidth}{!} {
    \begin{tabular}{l|c|c|c|c|c|c|c|c} 
    \toprule
       \multirow{2}{*}{\textbf{ Method}} & \multicolumn{4}{c|}{\textbf{Classification}} & \multicolumn{4}{c}{\textbf{Grouping}} \\ \cmidrule{2-9}
       &\textbf{Precision} & \textbf{Recall} & \textbf{F1-score} & \textbf{Accuracy}  & \textbf{Asso-prec.} & \textbf{Asso-rec.} & \textbf{IoU-prec.} & \textbf{IoU-rec.} \\
    \midrule
    Original Dataset & 0.916 &0.891 & 0.903 & 0.957& 0.818& 0.895& 0.764 &0.743  \\
    Figma Dataset & 0.882 &0.843 &0.862 & 0.928 & 0.775 & 0.855 & 0.724& 0.707 \\
    \bottomrule
    \end{tabular}
    }
    \label{tab:ablation_figma}
\end{table*}

\section{User study}
In this section, we conduct a user study to evaluate the effectiveness of applying our approach to Imgcook which is an automated code generation platform.

\subsection{Procedures}
This study recruited 10 developers, all of whom are proficient in UI development using the Vue framework, with an average of approximately three years of development experience. As described in previous sections, fragmented layers exist in UI icons, decorative components, and background components. For each category, we randomly pick five design prototypes and fetch corresponding components which all contain fragmented layers. We then generate front-end code by imgcook automatically. For the control group, developers evaluate and modify the front-end code of the original UI components containing fragmented layers. For the experimental group, the original design prototypes are processed and merged by our approach before code generation. The developers modify the front-end code to reach acceptable industry standards based on their own experience. Before the evaluation, each developer is given enough time to get familiar with the UI components. We use the git service to record the lines of code modified, and we also record the time of modifying front-end code by the developers. The fewer lines of code are modified, the higher the code availability is. So we evaluate the code availability by the formula:
\begin{equation}
    availability= 1-\frac{number\ of\ modified \ lines}{total\  number\  of \ lines}
\end{equation}
The developers do not know we record the time as the time pressure may affect their modification speed. When the developers finish the code modification for one UI component, they mark the generated code on a five-point Likert scale for readability and maintainability respectively (1: not readable and 5: strongly readable, so is maintainability). All developers evaluate and modify front-end code individually and they all do not know which code is merged by our approach. 

\begin{table*}[htp]
\centering
\caption{\textbf{Results of user study}} \label{tab:HumanEvaluation}
\resizebox{0.9\linewidth}{!}{
    \begin{tabular}{llccc} 
    \toprule
    \multicolumn{1}{c}{\multirow{2}{*}{Category}} & \multicolumn{1}{c}{\multirow{2}{*}{Measures}} & \multicolumn{3}{c}{Results} \\
    \cmidrule{3-5}
    \multicolumn{1}{c}{}  & \multicolumn{1}{c}{}  & \multicolumn{1}{c|}{control} & \multicolumn{1}{c|}{experimental} & \multicolumn{1}{c}{Delta}  \\ 
    \midrule
    \multirow{4}{*}{Icon}                        
    & availability  &$76.3$&$83.6^{**}$&+7.30 \\
    &  modification time (min) &$8.95$&$6.12^{*}\,$&$-$2.83\\
    & readability &$3.32$ &$4.40^{**}$&+1.08\\
    & maintainability  &$2.86$&$4.12^{**}$&+1.26\\ 
    \midrule
    \multirow{4}{*}{Decorative pattern}           
    & availability & $73.6$ & $81.5^{**}$&+7.90\\
    & modification time (min) & $9.24$ & $5.56^{**}$&$-$3.68\\
    & readability  &  $3.16$ & $4.08^{**}$&+0.92\\
    & maintainability &$2.82$&$3.74^{**}$&+0.92\\ 
    \midrule
    \multirow{4}{*}{Background}                   
    & availability &$75.2$ &$79.8^{*}\,$ &+4.60\\
    &  modification time (min) &$10.66$ &$8.34^{*}$&$-$2.32\\
    & readability &$2.06$ &$3.88^{**}$&+1.28\\
    & maintainability &$2.00$&$3.52^{**}$&+1.52\\ 
    \midrule
    \multirow{4}{*}{\textbf{Average}}                      
    & availability &$75.0$& $81.6$&+6.60\\
    & modification time (min)  &$9.61$ &$6.67$&$-$2.94\\
    & readability &$2.85$ &$4.12$&+1.27 \\
    & maintainability &$2.56$ &$3.79$&+1.23\\
    \bottomrule
    \end{tabular}
}
   \begin{tablenotes}
   \footnotesize
     \item[1] ** denotes $p < 0.01$ and * denotes $p < 0.05$
   \end{tablenotes}
\end{table*}

\subsection{Results}

\begin{figure*}[htb]
\centering 
\includegraphics[width=1\textwidth]{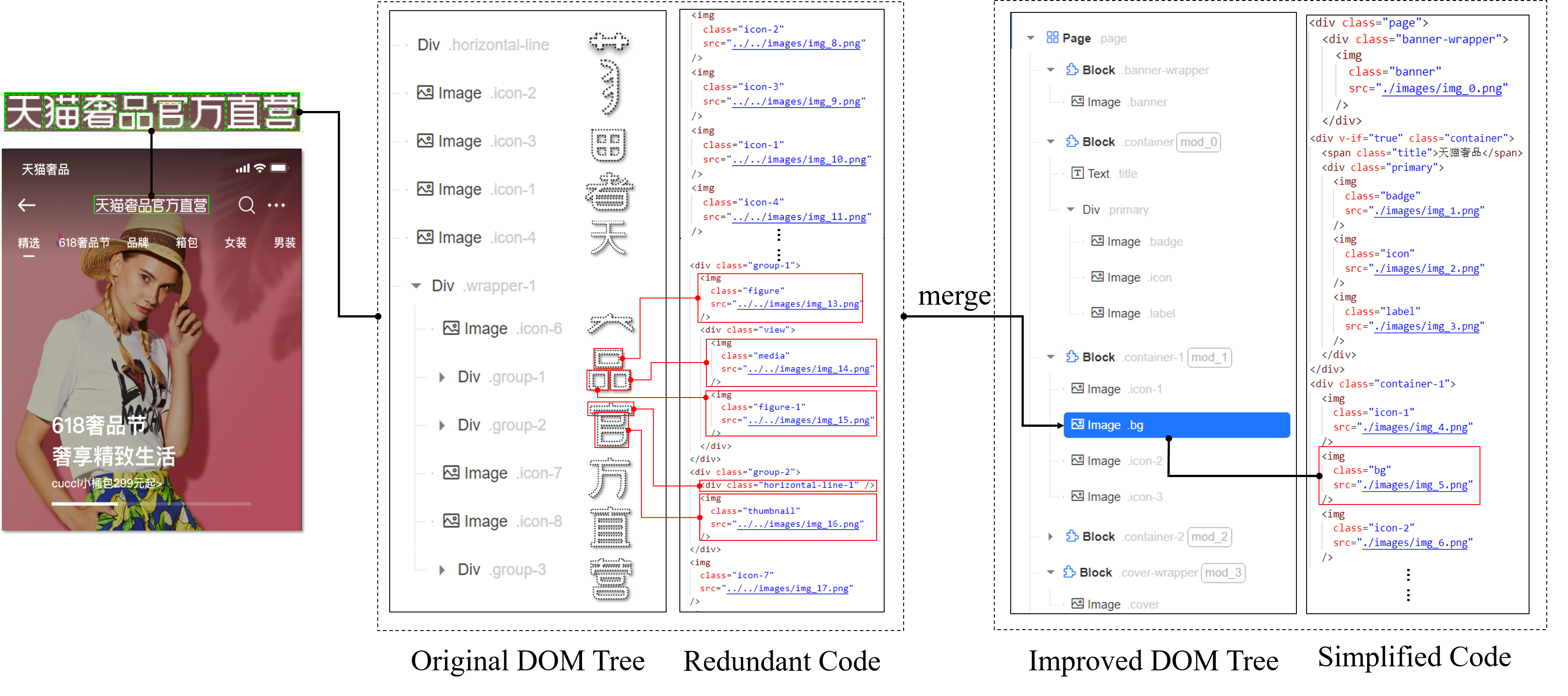} 
\caption{Results of improved DOM tree in imgcook after grouping fragmented layers by our approach.}
\label{codegeneration} 
\end{figure*}

Table \ref{tab:HumanEvaluation} shows that the number of code lines modified in the experimental group is less than the control group for all three categories. Furthermore, the time spent on modifying code to reach acceptable industry standards in the experimental group is also shorter than that in the control group. For example, the availability of code is improved by 7.3\% and 7.9\% for UI icons and decorative UI patterns. The average modification time is reduced by 42.1\% with our approach. It demonstrates that the availability of generated code is improved a lot by adopting our approach to merge fragmented layers in UI design prototypes before code generation. For readability and maintainability, Table \ref{tab:HumanEvaluation} shows that the developers generally mark higher scores for front-end code after merging fragmented layers. The average score of readability and maintainability in the experimental group is 44.6\% and 48.0\% more than that in the control group respectively. Above all, the generated code by imgcook has higher availability, readability, and maintainability after merging associative fragmented layers in design prototypes by our approach. 

To understand the significance of the difference, we also carry out the Mann-Whitney U test which is specifically designed for small samples. $p \leq 0.05$ is typically considered to be statistically significant and $p \leq 0.01$ is considered to be highly statistically significant. The results of the study show that our approach significantly outperforms the baselines in all four metrics. The average modification time decreased from 9.61 minutes in the control group to 6.67 minutes in the experimental group, resulting in a time savings of 2.94 minutes. We observed the statistical significance of time savings for three types of code modifications, indicating that the proposed method results in meaningful time savings compared to the baseline.

We also gather participants' feedback on the differences between the two versions of the code. Overall, most participants feel that the code generated after merging fragmented layers using our method is closer to a usable form in real-world business applications. On one hand, the total number of lines of code is generally reduced, as objects representing individual fragmented elements are merged. This reduction is particularly noticeable in decorative patterns, and the modification time differences further support this observation. Moreover, participants find the code produced by our method to be more consistent in terms of class naming and structure, making it easier to recognize and beneficial for future adjustments and reuse.

Finally, to further analyze the results above, Figure \ref{codegeneration} shows the results of an improved DOM tree after grouping fragmented layers. The DOM tree of the art font group has a complicated nested structure and many redundant image containers, which results in generating the wrong GUI running-time hierarchy and redundant code. Our method can group fragmented layers into a single group and add a "\#merge\#" tag for recognition by downstream code generation tools, such as imgcook. When imgcook recognizes the special tag, it merges all layers in the group to produce a single image container. After grouping fragmented layers, imgcook transforms art font into an image and uses one-line HTML code instead of interpreting fragmented layers as distinct entities. Overall, our method can facilitate imgcook to generate more maintainable and readable code.

\section{Conclusion}
In this study, we focus on grouping semantically consistent layers from original design files. Inspired by previous methods \cite{UILM,egfe}, we propose a new graph-learning model to classify UI layers and detect the boundary of merging groups. We demonstrate that our algorithm can not only avoid the disadvantages of previous methods, which are sample imbalance and failure at dealing with background groups but can combine their advantages, which are high accuracy of layer classification based on the multi-modal information and high quality of layer grouping due to accurate boundary regression. The experiments and a user study prove the effectiveness of our approach.\par
Our method abandons the UI screenshots which are useful for bounding box regression. In future work, we consider adopting the RoI align module \cite{he2017mask} to refine the box proposals of GNNs. By utilizing the global semantics of the whole UI images, we expect that our algorithm can be further improved.

\backmatter

\section*{Declarations}
\bmhead{Competing Interests}
The authors declare that they have no conflict of interest.
\bmhead{Data availability}
Dataset in this study is available at \url{https://zenodo.org/record/8022996}. Follow the guidelines on this website to download the dataset.
\bmhead{Code availability} 
Code of this study is available at \url{https://github.com/zjl12138/ULDGNN}

\bibliography{ulmgnn}


\begin{thebibliography}{53}
\ifx \bisbn   \undefined \def \bisbn  #1{ISBN #1}\fi
\ifx \binits  \undefined \def \binits#1{#1}\fi
\ifx \bauthor  \undefined \def \bauthor#1{#1}\fi
\ifx \batitle  \undefined \def \batitle#1{#1}\fi
\ifx \bjtitle  \undefined \def \bjtitle#1{#1}\fi
\ifx \bvolume  \undefined \def \bvolume#1{\textbf{#1}}\fi
\ifx \byear  \undefined \def \byear#1{#1}\fi
\ifx \bissue  \undefined \def \bissue#1{#1}\fi
\ifx \bfpage  \undefined \def \bfpage#1{#1}\fi
\ifx \blpage  \undefined \def \blpage #1{#1}\fi
\ifx \burl  \undefined \def \burl#1{\textsf{#1}}\fi
\ifx \doiurl  \undefined \def \doiurl#1{\url{https://doi.org/#1}}\fi
\ifx \betal  \undefined \def \betal{\textit{et al.}}\fi
\ifx \binstitute  \undefined \def \binstitute#1{#1}\fi
\ifx \binstitutionaled  \undefined \def \binstitutionaled#1{#1}\fi
\ifx \bctitle  \undefined \def \bctitle#1{#1}\fi
\ifx \beditor  \undefined \def \beditor#1{#1}\fi
\ifx \bpublisher  \undefined \def \bpublisher#1{#1}\fi
\ifx \bbtitle  \undefined \def \bbtitle#1{#1}\fi
\ifx \bedition  \undefined \def \bedition#1{#1}\fi
\ifx \bseriesno  \undefined \def \bseriesno#1{#1}\fi
\ifx \blocation  \undefined \def \blocation#1{#1}\fi
\ifx \bsertitle  \undefined \def \bsertitle#1{#1}\fi
\ifx \bsnm \undefined \def \bsnm#1{#1}\fi
\ifx \bsuffix \undefined \def \bsuffix#1{#1}\fi
\ifx \bparticle \undefined \def \bparticle#1{#1}\fi
\ifx \barticle \undefined \def \barticle#1{#1}\fi
\bibcommenthead
\ifx \bconfdate \undefined \def \bconfdate #1{#1}\fi
\ifx \botherref \undefined \def \botherref #1{#1}\fi
\ifx \url \undefined \def \url#1{\textsf{#1}}\fi
\ifx \bchapter \undefined \def \bchapter#1{#1}\fi
\ifx \bbook \undefined \def \bbook#1{#1}\fi
\ifx \bcomment \undefined \def \bcomment#1{#1}\fi
\ifx \oauthor \undefined \def \oauthor#1{#1}\fi
\ifx \citeauthoryear \undefined \def \citeauthoryear#1{#1}\fi
\ifx \endbibitem  \undefined \def \endbibitem {}\fi
\ifx \bconflocation  \undefined \def \bconflocation#1{#1}\fi
\ifx \arxivurl  \undefined \def \arxivurl#1{\textsf{#1}}\fi
\csname PreBibitemsHook\endcsname

\bibitem[\protect\citeauthoryear{}{2022}]{sketchcli}
\begin{botherref}
Sketch Command-line interface.
\url{https://developer.sketch.com/cli/}
(2022)
\end{botherref}
\endbibitem

\bibitem[\protect\citeauthoryear{}{2015}]{figma}
\begin{botherref}
Figma.
\url{https://www.figma.com/}
(2015)
\end{botherref}
\endbibitem

\bibitem[\protect\citeauthoryear{Moran et~al.}{2018}]{moran2018machine}
\begin{barticle}
\bauthor{\bsnm{Moran}, \binits{K.}},
\bauthor{\bsnm{Bernal-C{\'a}rdenas}, \binits{C.}},
\bauthor{\bsnm{Curcio}, \binits{M.}},
\bauthor{\bsnm{Bonett}, \binits{R.}},
\bauthor{\bsnm{Poshyvanyk}, \binits{D.}}:
\batitle{Machine learning-based prototyping of graphical user interfaces for mobile apps}.
\bjtitle{IEEE Transactions on Software Engineering}
\bvolume{46}(\bissue{2}),
\bfpage{196}--\blpage{221}
(\byear{2018})
\end{barticle}
\endbibitem

\bibitem[\protect\citeauthoryear{Alibaba}{2021}]{imgcook}
\begin{botherref}
\oauthor{\bsnm{Alibaba}}:
Intelligent Code Generation for Design Drafts
(2021).
\url{http://www.imgcook.com/ [Accessed on Dec. 27, 2021].}
\end{botherref}
\endbibitem

\bibitem[\protect\citeauthoryear{Beltramelli}{2018}]{beltramelli2018pix2code}
\begin{bchapter}
\bauthor{\bsnm{Beltramelli}, \binits{T.}}:
\bctitle{pix2code: Generating code from a graphical user interface screenshot}.
In: \bbtitle{Proceedings of the ACM SIGCHI Symposium on Engineering Interactive Computing Systems},
pp. \bfpage{1}--\blpage{6}
(\byear{2018})
\end{bchapter}
\endbibitem

\bibitem[\protect\citeauthoryear{Chen et~al.}{2018}]{GUI_skeleton}
\begin{bchapter}
\bauthor{\bsnm{Chen}, \binits{C.}},
\bauthor{\bsnm{Su}, \binits{T.}},
\bauthor{\bsnm{Meng}, \binits{G.}},
\bauthor{\bsnm{Xing}, \binits{Z.}},
\bauthor{\bsnm{Liu}, \binits{Y.}}:
\bctitle{From ui design image to gui skeleton: A neural machine translator to bootstrap mobile gui implementation}.
In: \bbtitle{2018 IEEE/ACM 40th International Conference on Software Engineering (ICSE)},
pp. \bfpage{665}--\blpage{676}
(\byear{2018}).
\doiurl{10.1145/3180155.3180240}
\end{bchapter}
\endbibitem

\bibitem[\protect\citeauthoryear{Mohian and Csallner}{2020}]{mohian2020doodle2app}
\begin{bchapter}
\bauthor{\bsnm{Mohian}, \binits{S.}},
\bauthor{\bsnm{Csallner}, \binits{C.}}:
\bctitle{Doodle2app: Native app code by freehand ui sketching}.
In: \bbtitle{Proceedings of the IEEE/ACM 7th International Conference on Mobile Software Engineering and Systems},
pp. \bfpage{81}--\blpage{84}
(\byear{2020})
\end{bchapter}
\endbibitem

\bibitem[\protect\citeauthoryear{Xiao et~al.}{2024}]{xiao2024prototype2code}
\begin{botherref}
\oauthor{\bsnm{Xiao}, \binits{S.}},
\oauthor{\bsnm{Chen}, \binits{Y.}},
\oauthor{\bsnm{Li}, \binits{J.}},
\oauthor{\bsnm{Chen}, \binits{L.}},
\oauthor{\bsnm{Sun}, \binits{L.}},
\oauthor{\bsnm{Zhou}, \binits{T.}}:
Prototype2code: End-to-end front-end code generation from ui design prototypes.
arXiv preprint arXiv:2405.04975
(2024)
\end{botherref}
\endbibitem

\bibitem[\protect\citeauthoryear{Yunnong et~al.}{2022}]{UILM}
\begin{botherref}
\oauthor{\bsnm{Yunnong}, \binits{C.}},
\oauthor{\bsnm{Yankun}, \binits{Z.}},
\oauthor{\bsnm{Chuning}, \binits{S.}},
\oauthor{\bsnm{Jiazhi}, \binits{L.}},
\oauthor{\bsnm{Liuqing}, \binits{C.}},
\oauthor{\bsnm{Zejian}, \binits{L.}},
\oauthor{\bsnm{Lingyun}, \binits{S.}},
\oauthor{\bsnm{Tingting}, \binits{Z.}},
\oauthor{\bsnm{Yanfang}, \binits{C.}}:
Ui layers merger: merging ui layers via visual learning and boundary prior.
Frontiers of Information Technology \& Electronic Engineering
(2022)
\end{botherref}
\endbibitem

\bibitem[\protect\citeauthoryear{Chen et~al.}{2024}]{egfe}
\begin{bchapter}
\bauthor{\bsnm{Chen}, \binits{L.}},
\bauthor{\bsnm{Chen}, \binits{Y.}},
\bauthor{\bsnm{Xiao}, \binits{S.}},
\bauthor{\bsnm{Song}, \binits{Y.}},
\bauthor{\bsnm{Sun}, \binits{L.}},
\bauthor{\bsnm{Zhen}, \binits{Y.}},
\bauthor{\bsnm{Zhou}, \binits{T.}},
\bauthor{\bsnm{Chang}, \binits{Y.}}:
\bctitle{Egfe: End-to-end grouping of fragmented elements in ui designs with multimodal learning}.
In: \bbtitle{Proceedings of the 46th IEEE/ACM International Conference on Software Engineering},
pp. \bfpage{1}--\blpage{12}
(\byear{2024})
\end{bchapter}
\endbibitem

\bibitem[\protect\citeauthoryear{Xie et~al.}{2022}]{UIED2}
\begin{bchapter}
\bauthor{\bsnm{Xie}, \binits{M.}},
\bauthor{\bsnm{Xing}, \binits{Z.}},
\bauthor{\bsnm{Feng}, \binits{S.}},
\bauthor{\bsnm{Xu}, \binits{X.}},
\bauthor{\bsnm{Zhu}, \binits{L.}},
\bauthor{\bsnm{Chen}, \binits{C.}}:
\bctitle{Psychologically-inspired, unsupervised inference of perceptual groups of gui widgets from gui images}.
In: \bbtitle{Proceedings of the 30th ACM Joint European Software Engineering Conference and Symposium on the Foundations of Software Engineering}.
\bsertitle{ESEC/FSE 2022},
pp. \bfpage{332}--\blpage{343}.
\bpublisher{Association for Computing Machinery},
\blocation{New York, NY, USA}
(\byear{2022}).
\doiurl{10.1145/3540250.3549138} .
\burl{https://doi.org/10.1145/3540250.3549138}
\end{bchapter}
\endbibitem

\bibitem[\protect\citeauthoryear{Li et~al.}{2022}]{li2022uldgnn}
\begin{botherref}
\oauthor{\bsnm{Li}, \binits{J.}},
\oauthor{\bsnm{Zhou}, \binits{T.}},
\oauthor{\bsnm{Chen}, \binits{Y.}},
\oauthor{\bsnm{Chang}, \binits{Y.}},
\oauthor{\bsnm{Zhen}, \binits{Y.}},
\oauthor{\bsnm{Sun}, \binits{L.}},
\oauthor{\bsnm{Chen}, \binits{L.}}:
ULDGNN: A Fragmented UI Layer Detector Based on Graph Neural Networks
(2022)
\end{botherref}
\endbibitem

\bibitem[\protect\citeauthoryear{Manandhar et~al.}{2021}]{manandhar2021magic}
\begin{bchapter}
\bauthor{\bsnm{Manandhar}, \binits{D.}},
\bauthor{\bsnm{Jin}, \binits{H.}},
\bauthor{\bsnm{Collomosse}, \binits{J.}}:
\bctitle{Magic layouts: Structural prior for component detection in user interface designs}.
In: \bbtitle{Proceedings of the IEEE/CVF Conference on Computer Vision and Pattern Recognition},
pp. \bfpage{15809}--\blpage{15818}
(\byear{2021})
\end{bchapter}
\endbibitem

\bibitem[\protect\citeauthoryear{He et~al.}{2021}]{he2021actionbert}
\begin{bchapter}
\bauthor{\bsnm{He}, \binits{Z.}},
\bauthor{\bsnm{Sunkara}, \binits{S.}},
\bauthor{\bsnm{Zang}, \binits{X.}},
\bauthor{\bsnm{Xu}, \binits{Y.}},
\bauthor{\bsnm{Liu}, \binits{L.}},
\bauthor{\bsnm{Wichers}, \binits{N.}},
\bauthor{\bsnm{Schubiner}, \binits{G.}},
\bauthor{\bsnm{Lee}, \binits{R.}},
\bauthor{\bsnm{Chen}, \binits{J.}}:
\bctitle{Actionbert: Leveraging user actions for semantic understanding of user interfaces}.
In: \bbtitle{Proceedings of the AAAI Conference on Artificial Intelligence},
vol. \bseriesno{35},
pp. \bfpage{5931}--\blpage{5938}
(\byear{2021})
\end{bchapter}
\endbibitem

\bibitem[\protect\citeauthoryear{Liu et~al.}{2018}]{liu2018learning}
\begin{bchapter}
\bauthor{\bsnm{Liu}, \binits{T.F.}},
\bauthor{\bsnm{Craft}, \binits{M.}},
\bauthor{\bsnm{Situ}, \binits{J.}},
\bauthor{\bsnm{Yumer}, \binits{E.}},
\bauthor{\bsnm{Mech}, \binits{R.}},
\bauthor{\bsnm{Kumar}, \binits{R.}}:
\bctitle{Learning design semantics for mobile apps}.
In: \bbtitle{Proceedings of the 31st Annual ACM Symposium on User Interface Software and Technology},
pp. \bfpage{569}--\blpage{579}
(\byear{2018})
\end{bchapter}
\endbibitem

\bibitem[\protect\citeauthoryear{Zang et~al.}{2021}]{multimodal_icon_annotation}
\begin{bchapter}
\bauthor{\bsnm{Zang}, \binits{X.}},
\bauthor{\bsnm{Xu}, \binits{Y.}},
\bauthor{\bsnm{Chen}, \binits{J.}}:
\bctitle{Multimodal icon annotation for mobile applications}.
In: \bbtitle{Proceedings of the 23rd International Conference on Mobile Human-Computer Interaction},
pp. \bfpage{1}--\blpage{11}
(\byear{2021})
\end{bchapter}
\endbibitem

\bibitem[\protect\citeauthoryear{Degott et~al.}{2019}]{degott2019learning}
\begin{bchapter}
\bauthor{\bsnm{Degott}, \binits{C.}},
\bauthor{\bsnm{Borges~Jr}, \binits{N.P.}},
\bauthor{\bsnm{Zeller}, \binits{A.}}:
\bctitle{Learning user interface element interactions}.
In: \bbtitle{Proceedings of the 28th ACM SIGSOFT International Symposium on Software Testing and Analysis},
pp. \bfpage{296}--\blpage{306}
(\byear{2019})
\end{bchapter}
\endbibitem

\bibitem[\protect\citeauthoryear{Li et~al.}{2019}]{humanoid}
\begin{bchapter}
\bauthor{\bsnm{Li}, \binits{Y.}},
\bauthor{\bsnm{Yang}, \binits{Z.}},
\bauthor{\bsnm{Guo}, \binits{Y.}},
\bauthor{\bsnm{Chen}, \binits{X.}}:
\bctitle{Humanoid: A deep learning-based approach to automated black-box android app testing}.
In: \bbtitle{2019 34th IEEE/ACM International Conference on Automated Software Engineering (ASE)},
pp. \bfpage{1070}--\blpage{1073}
(\byear{2019}).
\doiurl{10.1109/ASE.2019.00104}
\end{bchapter}
\endbibitem

\bibitem[\protect\citeauthoryear{Nguyen and Csallner}{2015}]{Revers_engi}
\begin{bchapter}
\bauthor{\bsnm{Nguyen}, \binits{T.A.}},
\bauthor{\bsnm{Csallner}, \binits{C.}}:
\bctitle{Reverse engineering mobile application user interfaces with remaui (t)}.
In: \bbtitle{2015 30th IEEE/ACM International Conference on Automated Software Engineering (ASE)},
pp. \bfpage{248}--\blpage{259}
(\byear{2015}).
\doiurl{10.1109/ASE.2015.32}
\end{bchapter}
\endbibitem

\bibitem[\protect\citeauthoryear{Zhang et~al.}{2021}]{screen_recog}
\begin{bchapter}
\bauthor{\bsnm{Zhang}, \binits{X.}},
\bauthor{\bsnm{Greef}, \binits{L.}},
\bauthor{\bsnm{Swearngin}, \binits{A.}},
\bauthor{\bsnm{White}, \binits{S.}},
\bauthor{\bsnm{Murray}, \binits{K.}},
\bauthor{\bsnm{Yu}, \binits{L.}},
\bauthor{\bsnm{Shan}, \binits{Q.}},
\bauthor{\bsnm{Nichols}, \binits{J.}},
\bauthor{\bsnm{Wu}, \binits{J.}},
\bauthor{\bsnm{Fleizach}, \binits{C.}},
\bauthor{\bsnm{Everitt}, \binits{A.}},
\bauthor{\bsnm{Bigham}, \binits{J.P.}}:
\bctitle{Screen recognition: Creating accessibility metadata for mobile applications from pixels}.
In: \bbtitle{Proceedings of the 2021 CHI Conference on Human Factors in Computing Systems}.
\bsertitle{CHI '21}.
\bpublisher{Association for Computing Machinery},
\blocation{New York, NY, USA}
(\byear{2021}).
\doiurl{10.1145/3411764.3445186} .
\burl{https://doi.org/10.1145/3411764.3445186}
\end{bchapter}
\endbibitem

\bibitem[\protect\citeauthoryear{Xiao et~al.}{2022}]{xiao2022ui}
\begin{bchapter}
\bauthor{\bsnm{Xiao}, \binits{S.}},
\bauthor{\bsnm{Zhou}, \binits{T.}},
\bauthor{\bsnm{Chen}, \binits{Y.}},
\bauthor{\bsnm{Zhang}, \binits{D.}},
\bauthor{\bsnm{Chen}, \binits{L.}},
\bauthor{\bsnm{Sun}, \binits{L.}},
\bauthor{\bsnm{Yue}, \binits{S.}}:
\bctitle{Ui layers group detector: Grouping ui layers via text fusion and box attention}.
In: \bbtitle{CAAI International Conference on Artificial Intelligence},
pp. \bfpage{303}--\blpage{314}
(\byear{2022}).
\bcomment{Springer}
\end{bchapter}
\endbibitem

\bibitem[\protect\citeauthoryear{Xiao et~al.}{2024}]{xiao2024ui}
\begin{barticle}
\bauthor{\bsnm{Xiao}, \binits{S.}},
\bauthor{\bsnm{Chen}, \binits{Y.}},
\bauthor{\bsnm{Song}, \binits{Y.}},
\bauthor{\bsnm{Chen}, \binits{L.}},
\bauthor{\bsnm{Sun}, \binits{L.}},
\bauthor{\bsnm{Zhen}, \binits{Y.}},
\bauthor{\bsnm{Chang}, \binits{Y.}},
\bauthor{\bsnm{Zhou}, \binits{T.}}:
\batitle{Ui semantic component group detection: Grouping ui elements with similar semantics in mobile graphical user interface}.
\bjtitle{Displays}
\bvolume{83},
\bfpage{102679}
(\byear{2024})
\end{barticle}
\endbibitem

\bibitem[\protect\citeauthoryear{Zheng et~al.}{2019}]{faceoff}
\begin{bchapter}
\bauthor{\bsnm{Zheng}, \binits{S.}},
\bauthor{\bsnm{Hu}, \binits{Z.}},
\bauthor{\bsnm{Ma}, \binits{Y.}}:
\bctitle{Faceoff: Assisting the manifestation design of web graphical user interface}.
In: \bbtitle{Proceedings of the Twelfth ACM International Conference on Web Search and Data Mining}.
\bsertitle{WSDM '19},
pp. \bfpage{774}--\blpage{777}.
\bpublisher{Association for Computing Machinery},
\blocation{New York, NY, USA}
(\byear{2019}).
\doiurl{10.1145/3289600.3290610} .
\burl{https://doi.org/10.1145/3289600.3290610}
\end{bchapter}
\endbibitem

\bibitem[\protect\citeauthoryear{Bajammal et~al.}{2018}]{gen_reu_webcomp}
\begin{bchapter}
\bauthor{\bsnm{Bajammal}, \binits{M.}},
\bauthor{\bsnm{Mazinanian}, \binits{D.}},
\bauthor{\bsnm{Mesbah}, \binits{A.}}:
\bctitle{Generating reusable web components from mockups}.
In: \bbtitle{2018 33rd IEEE/ACM International Conference on Automated Software Engineering (ASE)},
pp. \bfpage{601}--\blpage{611}
(\byear{2018}).
\doiurl{10.1145/3238147.3238194}
\end{bchapter}
\endbibitem

\bibitem[\protect\citeauthoryear{Bielik et~al.}{2018}]{Robust_Relational_Layout}
\begin{botherref}
\oauthor{\bsnm{Bielik}, \binits{P.}},
\oauthor{\bsnm{Fischer}, \binits{M.}},
\oauthor{\bsnm{Vechev}, \binits{M.}}:
Robust relational layout synthesis from examples for android.
Proc. ACM Program. Lang.
\textbf{2}(OOPSLA)
(2018)
\doiurl{10.1145/3276526}
\end{botherref}
\endbibitem

\bibitem[\protect\citeauthoryear{Yandrapally et~al.}{2020}]{Near-Duplicate}
\begin{bchapter}
\bauthor{\bsnm{Yandrapally}, \binits{R.}},
\bauthor{\bsnm{Stocco}, \binits{A.}},
\bauthor{\bsnm{Mesbah}, \binits{A.}}:
\bctitle{Near-duplicate detection in web app model inference}.
In: \bbtitle{Proceedings of the ACM/IEEE 42nd International Conference on Software Engineering}.
\bsertitle{ICSE '20},
pp. \bfpage{186}--\blpage{197}.
\bpublisher{Association for Computing Machinery},
\blocation{New York, NY, USA}
(\byear{2020}).
\doiurl{10.1145/3377811.3380416} .
\burl{https://doi.org/10.1145/3377811.3380416}
\end{bchapter}
\endbibitem

\bibitem[\protect\citeauthoryear{Gori et~al.}{2005}]{gori2005new}
\begin{bchapter}
\bauthor{\bsnm{Gori}, \binits{M.}},
\bauthor{\bsnm{Monfardini}, \binits{G.}},
\bauthor{\bsnm{Scarselli}, \binits{F.}}:
\bctitle{A new model for learning in graph domains}.
In: \bbtitle{Proceedings. 2005 IEEE International Joint Conference on Neural Networks, 2005.},
vol. \bseriesno{2},
pp. \bfpage{729}--\blpage{734}
(\byear{2005}).
\bcomment{IEEE}
\end{bchapter}
\endbibitem

\bibitem[\protect\citeauthoryear{Bruna et~al.}{2014}]{bruna6203spectral}
\begin{bchapter}
\bauthor{\bsnm{Bruna}, \binits{J.}},
\bauthor{\bsnm{Zaremba}, \binits{W.}},
\bauthor{\bsnm{Szlam}, \binits{A.}},
\bauthor{\bsnm{LeCun}, \binits{Y.}}:
\bctitle{Spectral networks and deep locally connected networks on graphs}.
In: \bbtitle{2nd International Conference on Learning Representations, ICLR 2014}
(\byear{2014})
\end{bchapter}
\endbibitem

\bibitem[\protect\citeauthoryear{Defferrard et~al.}{2016}]{defferrard2016convolutionalneuralNetworkongraphs}
\begin{barticle}
\bauthor{\bsnm{Defferrard}, \binits{M.}},
\bauthor{\bsnm{Bresson}, \binits{X.}},
\bauthor{\bsnm{Vandergheynst}, \binits{P.}}:
\batitle{Convolutional neural networks on graphs with fast localized spectral filtering}.
\bjtitle{Advances in neural information processing systems}
\bvolume{29},
\bfpage{3844}--\blpage{3852}
(\byear{2016})
\end{barticle}
\endbibitem

\bibitem[\protect\citeauthoryear{Kipf and Welling}{2022}]{kipf2022semi}
\begin{bchapter}
\bauthor{\bsnm{Kipf}, \binits{T.N.}},
\bauthor{\bsnm{Welling}, \binits{M.}}:
\bctitle{Semi-supervised classification with graph convolutional networks}.
In: \bbtitle{International Conference on Learning Representations}
(\byear{2022})
\end{bchapter}
\endbibitem

\bibitem[\protect\citeauthoryear{Monti et~al.}{2017}]{monti2017geometric}
\begin{bchapter}
\bauthor{\bsnm{Monti}, \binits{F.}},
\bauthor{\bsnm{Boscaini}, \binits{D.}},
\bauthor{\bsnm{Masci}, \binits{J.}},
\bauthor{\bsnm{Rodola}, \binits{E.}},
\bauthor{\bsnm{Svoboda}, \binits{J.}},
\bauthor{\bsnm{Bronstein}, \binits{M.M.}}:
\bctitle{Geometric deep learning on graphs and manifolds using mixture model cnns}.
In: \bbtitle{Proceedings of the IEEE Conference on Computer Vision and Pattern Recognition},
pp. \bfpage{5115}--\blpage{5124}
(\byear{2017})
\end{bchapter}
\endbibitem

\bibitem[\protect\citeauthoryear{Hamilton et~al.}{2017}]{GraphSAGE}
\begin{bchapter}
\bauthor{\bsnm{Hamilton}, \binits{W.L.}},
\bauthor{\bsnm{Ying}, \binits{R.}},
\bauthor{\bsnm{Leskovec}, \binits{J.}}:
\bctitle{Inductive representation learning on large graphs}.
In: \bbtitle{Proceedings of the 31st International Conference on Neural Information Processing Systems},
pp. \bfpage{1025}--\blpage{1035}
(\byear{2017})
\end{bchapter}
\endbibitem

\bibitem[\protect\citeauthoryear{Veli{\v{c}}kovi{\'c} et~al.}{}]{velivckovicgraph}
\begin{botherref}
\oauthor{\bsnm{Veli{\v{c}}kovi{\'c}}, \binits{P.}},
\oauthor{\bsnm{Cucurull}, \binits{G.}},
\oauthor{\bsnm{Casanova}, \binits{A.}},
\oauthor{\bsnm{Romero}, \binits{A.}},
\oauthor{\bsnm{Li{\`o}}, \binits{P.}},
\oauthor{\bsnm{Bengio}, \binits{Y.}}:
Graph attention networks.
In: International Conference on Learning Representations
\end{botherref}
\endbibitem

\bibitem[\protect\citeauthoryear{Brody et~al.}{}]{brodyattentive}
\begin{botherref}
\oauthor{\bsnm{Brody}, \binits{S.}},
\oauthor{\bsnm{Alon}, \binits{U.}},
\oauthor{\bsnm{Yahav}, \binits{E.}}:
How attentive are graph attention networks?
In: International Conference on Learning Representations
\end{botherref}
\endbibitem

\bibitem[\protect\citeauthoryear{Ruiz et~al.}{2020}]{ruiz2020gated}
\begin{barticle}
\bauthor{\bsnm{Ruiz}, \binits{L.}},
\bauthor{\bsnm{Gama}, \binits{F.}},
\bauthor{\bsnm{Ribeiro}, \binits{A.}}:
\batitle{Gated graph recurrent neural networks}.
\bjtitle{IEEE Transactions on Signal Processing}
\bvolume{68},
\bfpage{6303}--\blpage{6318}
(\byear{2020})
\end{barticle}
\endbibitem

\bibitem[\protect\citeauthoryear{Xu et~al.}{2018}]{xu2018GIN}
\begin{bchapter}
\bauthor{\bsnm{Xu}, \binits{K.}},
\bauthor{\bsnm{Hu}, \binits{W.}},
\bauthor{\bsnm{Leskovec}, \binits{J.}},
\bauthor{\bsnm{Jegelka}, \binits{S.}}:
\bctitle{How powerful are graph neural networks?}
In: \bbtitle{International Conference on Learning Representations}
(\byear{2018})
\end{bchapter}
\endbibitem

\bibitem[\protect\citeauthoryear{Li et~al.}{2019}]{li2019deepgcns}
\begin{bchapter}
\bauthor{\bsnm{Li}, \binits{G.}},
\bauthor{\bsnm{Muller}, \binits{M.}},
\bauthor{\bsnm{Thabet}, \binits{A.}},
\bauthor{\bsnm{Ghanem}, \binits{B.}}:
\bctitle{Deepgcns: Can gcns go as deep as cnns?}
In: \bbtitle{Proceedings of the IEEE/CVF International Conference on Computer Vision},
pp. \bfpage{9267}--\blpage{9276}
(\byear{2019})
\end{bchapter}
\endbibitem

\bibitem[\protect\citeauthoryear{Chen et~al.}{2020}]{chen2020GCNII}
\begin{bchapter}
\bauthor{\bsnm{Chen}, \binits{M.}},
\bauthor{\bsnm{Wei}, \binits{Z.}},
\bauthor{\bsnm{Huang}, \binits{Z.}},
\bauthor{\bsnm{Ding}, \binits{B.}},
\bauthor{\bsnm{Li}, \binits{Y.}}:
\bctitle{Simple and deep graph convolutional networks}.
In: \bbtitle{International Conference on Machine Learning},
pp. \bfpage{1725}--\blpage{1735}
(\byear{2020}).
\bcomment{PMLR}
\end{bchapter}
\endbibitem

\bibitem[\protect\citeauthoryear{Ramp{\'a}{\v{s}}ek et~al.}{2022}]{rampavsek2022recipe}
\begin{barticle}
\bauthor{\bsnm{Ramp{\'a}{\v{s}}ek}, \binits{L.}},
\bauthor{\bsnm{Galkin}, \binits{M.}},
\bauthor{\bsnm{Dwivedi}, \binits{V.P.}},
\bauthor{\bsnm{Luu}, \binits{A.T.}},
\bauthor{\bsnm{Wolf}, \binits{G.}},
\bauthor{\bsnm{Beaini}, \binits{D.}}:
\batitle{Recipe for a general, powerful, scalable graph transformer}.
\bjtitle{Advances in Neural Information Processing Systems}
\bvolume{35},
\bfpage{14501}--\blpage{14515}
(\byear{2022})
\end{barticle}
\endbibitem

\bibitem[\protect\citeauthoryear{Ying et~al.}{2021}]{ying2021transformers}
\begin{barticle}
\bauthor{\bsnm{Ying}, \binits{C.}},
\bauthor{\bsnm{Cai}, \binits{T.}},
\bauthor{\bsnm{Luo}, \binits{S.}},
\bauthor{\bsnm{Zheng}, \binits{S.}},
\bauthor{\bsnm{Ke}, \binits{G.}},
\bauthor{\bsnm{He}, \binits{D.}},
\bauthor{\bsnm{Shen}, \binits{Y.}},
\bauthor{\bsnm{Liu}, \binits{T.-Y.}}:
\batitle{Do transformers really perform badly for graph representation?}
\bjtitle{Advances in Neural Information Processing Systems}
\bvolume{34},
\bfpage{28877}--\blpage{28888}
(\byear{2021})
\end{barticle}
\endbibitem

\bibitem[\protect\citeauthoryear{Shi and Rajkumar}{2020}]{shi2020pointGNN}
\begin{bchapter}
\bauthor{\bsnm{Shi}, \binits{W.}},
\bauthor{\bsnm{Rajkumar}, \binits{R.}}:
\bctitle{Point-gnn: Graph neural network for 3d object detection in a point cloud}.
In: \bbtitle{Proceedings of the IEEE/CVF Conference on Computer Vision and Pattern Recognition},
pp. \bfpage{1711}--\blpage{1719}
(\byear{2020})
\end{bchapter}
\endbibitem

\bibitem[\protect\citeauthoryear{Wen et~al.}{2019}]{wen2019graphskeleton}
\begin{bchapter}
\bauthor{\bsnm{Wen}, \binits{Y.-H.}},
\bauthor{\bsnm{Gao}, \binits{L.}},
\bauthor{\bsnm{Fu}, \binits{H.}},
\bauthor{\bsnm{Zhang}, \binits{F.-L.}},
\bauthor{\bsnm{Xia}, \binits{S.}}:
\bctitle{Graph cnns with motif and variable temporal block for skeleton-based action recognition}.
In: \bbtitle{Proceedings of the AAAI Conference on Artificial Intelligence},
vol. \bseriesno{33},
pp. \bfpage{8989}--\blpage{8996}
(\byear{2019})
\end{bchapter}
\endbibitem

\bibitem[\protect\citeauthoryear{Qi et~al.}{2017}]{qi20173d}
\begin{bchapter}
\bauthor{\bsnm{Qi}, \binits{X.}},
\bauthor{\bsnm{Liao}, \binits{R.}},
\bauthor{\bsnm{Jia}, \binits{J.}},
\bauthor{\bsnm{Fidler}, \binits{S.}},
\bauthor{\bsnm{Urtasun}, \binits{R.}}:
\bctitle{3d graph neural networks for rgbd semantic segmentation}.
In: \bbtitle{Proceedings of the IEEE International Conference on Computer Vision},
pp. \bfpage{5199}--\blpage{5208}
(\byear{2017})
\end{bchapter}
\endbibitem

\bibitem[\protect\citeauthoryear{Ang and Lim}{2022}]{HAMP}
\begin{bchapter}
\bauthor{\bsnm{Ang}, \binits{G.}},
\bauthor{\bsnm{Lim}, \binits{E.P.}}:
\bctitle{Learning user interface semantics from heterogeneous networks with multimodal and positional attributes}.
In: \bbtitle{27th International Conference on Intelligent User Interfaces}.
\bsertitle{IUI '22},
pp. \bfpage{433}--\blpage{446}.
\bpublisher{Association for Computing Machinery},
\blocation{New York, NY, USA}
(\byear{2022}).
\doiurl{10.1145/3490099.3511143} .
\burl{https://doi.org/10.1145/3490099.3511143}
\end{bchapter}
\endbibitem

\bibitem[\protect\citeauthoryear{Li et~al.}{2022}]{layout_dinoiser}
\begin{bchapter}
\bauthor{\bsnm{Li}, \binits{G.}},
\bauthor{\bsnm{Baechler}, \binits{G.}},
\bauthor{\bsnm{Tragut}, \binits{M.}},
\bauthor{\bsnm{Li}, \binits{Y.}}:
\bctitle{Learning to denoise raw mobile ui layouts for improving datasets at scale}.
In: \bbtitle{Proceedings of the 2022 CHI Conference on Human Factors in Computing Systems}.
\bsertitle{CHI '22}.
\bpublisher{Association for Computing Machinery},
\blocation{New York, NY, USA}
(\byear{2022}).
\doiurl{10.1145/3491102.3502042} .
\burl{https://doi.org/10.1145/3491102.3502042}
\end{bchapter}
\endbibitem

\bibitem[\protect\citeauthoryear{Mildenhall et~al.}{2020}]{mildenhall2020nerf}
\begin{bchapter}
\bauthor{\bsnm{Mildenhall}, \binits{B.}},
\bauthor{\bsnm{Srinivasan}, \binits{P.P.}},
\bauthor{\bsnm{Tancik}, \binits{M.}},
\bauthor{\bsnm{Barron}, \binits{J.T.}},
\bauthor{\bsnm{Ramamoorthi}, \binits{R.}},
\bauthor{\bsnm{Ng}, \binits{R.}}:
\bctitle{Nerf: Representing scenes as neural radiance fields for view synthesis}.
In: \bbtitle{Computer Vision--ECCV 2020: 16th European Conference, Glasgow, UK, August 23--28, 2020, Proceedings, Part I},
pp. \bfpage{405}--\blpage{421}
(\byear{2020})
\end{bchapter}
\endbibitem

\bibitem[\protect\citeauthoryear{Girshick}{2015}]{girshick2015fast}
\begin{bchapter}
\bauthor{\bsnm{Girshick}, \binits{R.}}:
\bctitle{Fast r-cnn}.
In: \bbtitle{Proceedings of the IEEE International Conference on Computer Vision},
pp. \bfpage{1440}--\blpage{1448}
(\byear{2015})
\end{bchapter}
\endbibitem

\bibitem[\protect\citeauthoryear{Ross and Doll{\'a}r}{2017}]{ross2017focal}
\begin{bchapter}
\bauthor{\bsnm{Ross}, \binits{T.-Y.}},
\bauthor{\bsnm{Doll{\'a}r}, \binits{G.}}:
\bctitle{Focal loss for dense object detection}.
In: \bbtitle{Proceedings of the IEEE Conference on Computer Vision and Pattern Recognition},
pp. \bfpage{2980}--\blpage{2988}
(\byear{2017})
\end{bchapter}
\endbibitem

\bibitem[\protect\citeauthoryear{Zheng et~al.}{2020}]{zheng2020distance}
\begin{bchapter}
\bauthor{\bsnm{Zheng}, \binits{Z.}},
\bauthor{\bsnm{Wang}, \binits{P.}},
\bauthor{\bsnm{Liu}, \binits{W.}},
\bauthor{\bsnm{Li}, \binits{J.}},
\bauthor{\bsnm{Ye}, \binits{R.}},
\bauthor{\bsnm{Ren}, \binits{D.}}:
\bctitle{Distance-iou loss: Faster and better learning for bounding box regression}.
In: \bbtitle{Proceedings of the AAAI Conference on Artificial Intelligence},
vol. \bseriesno{34},
pp. \bfpage{12993}--\blpage{13000}
(\byear{2020})
\end{bchapter}
\endbibitem

\bibitem[\protect\citeauthoryear{He et~al.}{2016}]{he2016deep}
\begin{bchapter}
\bauthor{\bsnm{He}, \binits{K.}},
\bauthor{\bsnm{Zhang}, \binits{X.}},
\bauthor{\bsnm{Ren}, \binits{S.}},
\bauthor{\bsnm{Sun}, \binits{J.}}:
\bctitle{Deep residual learning for image recognition}.
In: \bbtitle{Proceedings of the IEEE Conference on Computer Vision and Pattern Recognition},
pp. \bfpage{770}--\blpage{778}
(\byear{2016})
\end{bchapter}
\endbibitem

\bibitem[\protect\citeauthoryear{Hu et~al.}{}]{hustrategies}
\begin{botherref}
\oauthor{\bsnm{Hu}, \binits{W.}},
\oauthor{\bsnm{Liu}, \binits{B.}},
\oauthor{\bsnm{Gomes}, \binits{J.}},
\oauthor{\bsnm{Zitnik}, \binits{M.}},
\oauthor{\bsnm{Liang}, \binits{P.}},
\oauthor{\bsnm{Pande}, \binits{V.}},
\oauthor{\bsnm{Leskovec}, \binits{J.}}:
Strategies for pre-training graph neural networks.
In: International Conference on Learning Representations
\end{botherref}
\endbibitem

\bibitem[\protect\citeauthoryear{Loshchilov and Hutter}{}]{AdamW}
\begin{botherref}
\oauthor{\bsnm{Loshchilov}, \binits{I.}},
\oauthor{\bsnm{Hutter}, \binits{F.}}:
Decoupled weight decay regularization.
In: International Conference on Learning Representations
\end{botherref}
\endbibitem

\bibitem[\protect\citeauthoryear{He et~al.}{2017}]{he2017mask}
\begin{bchapter}
\bauthor{\bsnm{He}, \binits{K.}},
\bauthor{\bsnm{Gkioxari}, \binits{G.}},
\bauthor{\bsnm{Doll{\'a}r}, \binits{P.}},
\bauthor{\bsnm{Girshick}, \binits{R.}}:
\bctitle{Mask r-cnn}.
In: \bbtitle{Proceedings of the IEEE International Conference on Computer Vision},
pp. \bfpage{2961}--\blpage{2969}
(\byear{2017})
\end{bchapter}
\endbibitem

\end{thebibliography}

\end{document}